\documentclass[10pt]{iopart}
\usepackage{graphicx,epsfig}
\usepackage{bm}
\usepackage{iopams}
\usepackage{pdfsync}
\usepackage{color,ulem}

\newcommand{\ba}{\begin{eqnarray}}
\newcommand{\ea}{\end{eqnarray}}
\newcommand{\bse}{\numparts}
\newcommand{\ese}{\endnumparts}

\newcommand{\HH}{{\cal{H}}}
\newcommand{\KK}{{\cal{K}}}
\newcommand{\PP}{{\cal{P}}}
\newcommand{\RR}{{}^{(3)}{\cal{R}}}

\newcommand{\dd}{{\rm{d}}}
\newcommand{\bdd}{{\textrm{\bf{d}}}}
\newcommand{\eigf}{{\textrm{\bf{e}}}}

\newcommand{\Tgr}{T_{\tiny{\textrm{gr}}}}
\newcommand{\sgr}{s_{\tiny{\textrm{gr}}}}
\newcommand{\sgrls}{s_{\tiny{\textrm{grLS}}}}
\newcommand{\dotsgr}{\dot s_{\tiny{\textrm{gr}}}}

\newcommand{\pgr}{p_{\tiny{\textrm{gr}}}}
\newcommand{\qgr}{q^a_{\tiny{\textrm{gr}}}}
\newcommand{\Pigr}{\Pi^{ab}_{\tiny{\textrm{gr}}}}
\newcommand{\rhogr}{\rho_{\tiny{\textrm{gr}}}}
\newcommand{\SSgr}{S_{\tiny{\textrm{gr}}}}

\newcommand{\Da}{\delta^{(A)}}
\newcommand{\DDa}{\textrm{D}^{(A)}}

\newcommand{\DDrho}{\textrm{D}^{(\rho)}}

\newcommand{\DDmu}{\textrm{D}^{(\mu)}}
\newcommand{\DDH}{\textrm{D}^{(\HH)}}
\newcommand{\DDh}{\textrm{D}^{(h)}}

\newcommand{\DHH}{\delta^{(\HH)}}

\newcommand{\DDKK}{\textrm{D}^{(\KK)}}

\newcommand{\OmmO}{\bar\Omega_0 ^{(m)}}

\newcommand{\OmLO}{\bar\Omega_0 ^{(\Lambda)}}

\newcommand{\tbb}{t_{\textrm{\tiny{bb}}}}

\newcommand{\tls}{t_{\textrm{\tiny{LS}}}}
\newcommand{\als}{a_{\textrm{\tiny{LS}}}}
\newcommand{\Gals}{\Gamma_{\textrm{\tiny{LS}}}}

\newcommand{\hqls}{h_{q\textrm{\tiny{LS}}}}
\newcommand{\muqls}{\mu_{q\textrm{\tiny{LS}}}}
\newcommand{\muls}{\mu_{\textrm{\tiny{LS}}}}
\newcommand{\kqls}{\kappa_{q\textrm{\tiny{LS}}}}
\newcommand{\kls}{\kappa_{\textrm{\tiny{LS}}}}

\newcommand{\rhols}{\rho_{\textrm{\tiny{LS}}}}
\newcommand{\rhoqls}{\rho_{q\textrm{\tiny{LS}}}}
\newcommand{\HHqls}{{\cal {H}}_{q\textrm{\tiny{LS}}}}

\newcommand{\KKqls}{{\cal K}_{q\textrm{\tiny{LS}}}}
\newcommand{\KKls}{{\cal K}_{\textrm{\tiny{LS}}}}

\newcommand{\Omqmls}{\Omega_{q\textrm{\tiny{LS}}}^{(m)}}
\newcommand{\OmqLls}{\Omega_{q\textrm{\tiny{LS}}}^{(\Lambda)}}
\newcommand{\Omqkls}{\Omega_{q\textrm{\tiny{LS}}}^{(k)}}

\newcommand{\DDrhols}{\textrm{D}^{(\rho)}_{\textrm{\tiny{LS}}}}
\newcommand{\DHls}{\delta_{\textrm{\tiny{LS}}}^{(\HH)}}

\newcommand{\DDKKls}{\textrm{D}_{\textrm{\tiny{LS}}}^{(\KK)}}

\newcommand{\Dmuls}{\delta^{(\mu)}_{\textrm{\tiny{LS}}}}
\newcommand{\DDmuls}{\textrm{D}^{(\mu)}_{\textrm{\tiny{LS}}}}
\newcommand{\Dhls}{\delta_{\textrm{\tiny{LS}}}^{(h)}}
\newcommand{\DDhls}{\textrm{D}_{\textrm{\tiny{LS}}}^{(h)}}
\newcommand{\Dkls}{\delta_{\textrm{\tiny{LS}}}^{(\kappa)}}
\newcommand{\DDkls}{\textrm{D}_{\textrm{\tiny{LS}}}^{(\kappa)}}

\newcommand{\Hls}{\bar H_{\textrm{\tiny{LS}}}}
\newcommand{\Ommls}{\bar\Omega_{\textrm{\tiny{LS}}}^{(m)}}
\newcommand{\OmLls}{\bar\Omega_{\textrm{\tiny{LS}}}^{(\Lambda)}}
\newcommand{\Omkls}{\bar\Omega_{\textrm{\tiny{LS}}}^{(k)}}

\newcommand{\DDHas}{\textrm{D}_{\textrm{\tiny{as}}}^{(\HH)}}

\newcommand{\Drhoas}{\delta_{\textrm{\tiny{as}}}^{(\rho)}}
\newcommand{\rhobaras}{\bar\rho_{\textrm{\tiny{as}}}}
\newcommand{\HHbaras}{\bar\HH_{\textrm{\tiny{as}}}}

\begin{document}


\title[Gravitational entropy of local cosmic voids.]{Gravitational entropy of local cosmic voids.}
\author{Roberto A. Sussman$^{1}$ and Julien Larena$^{2}$} 
\address{$^{1}$Instituto de Ciencias Nucleares, Universidad Nacional Aut\'onoma de M\'exico (ICN-UNAM). A. P. 70--543, 04510 M\'exico D. F.\\
$^{2}$Department of Mathematics, Rhodes University, Grahamstown 6140, South Africa}
\eads{$^{1}$\mailto{sussman@nucleares.unam.mx}, $^{2}$\mailto{j.larena@ru.ac.za}}
%
%
\date{\today}
\begin{abstract}
We undertake a non--perturbative study of the evolution of the ``gravitational entropy'' proposed by Clifton, Ellis and Tavakol (CET) on local expanding cosmic CDM voids of $\sim 50-100$ Mpc size described as spherical under--dense regions with negative spatial curvature, whose dynamics is determined by Lema\^\i tre--Tolman--Bondi (LTB) dust models asymptotic to three different types of FLRW background: $\Lambda$CDM, Einstein de Sitter and ``open'' FLRW with $\Lambda=0$ and negative spatial curvature. By assuming generic nearly spatially flat and linear initial conditions at the last scattering time, we examine analytically and numerically the CET entropy evolution into a fully non--linear regime in our present cosmic time and beyond. Both analytic and numerical analysis reveal that the late time CET entropy growth is determined by the amplitude of initial fluctuations of spatial curvature at the last scattering time. This entropy growth decays to zero in the late asymptotic time range for all voids, but at a faster rate in voids with $\Lambda$CDM and open FLRW backgrounds. However, only for voids in a $\Lambda$CDM  background this suppression is sufficiently rapid for the CET entropy itself to reach a terminal equilibrium (or ``saturation'') value. The CET gravitational temperature vanishes asymptotically if $\Lambda=0$ and becomes asymptotically  proportional to $\Lambda$ for voids in a $\Lambda$CDM  background. In the linear regime of the LTB evolution our results coincide, qualitatively and quantitatively, with previous results based on linear perturbation theory.     
\end{abstract}

\maketitle
\section{Introduction}\label{intro}

One of the long standing open  issues in current theoretical physics is the proper definition of a ``gravitational'' entropy, $\sgr$, that determines the directionality of the gravitational interaction at all scales and energies. Hence, it is a different (though possibly related) notion from that of the thermal entropy of the field sources (hydrodynamical or non--collisional) or the holographic black hole entropies. Research on this issue has produced various self--consistent proposals, from Penrose's old idea of the ``arrow of time'' \cite{Penrose} (see also \cite{arrow}), to the proposals by  Clifton, Ellis and Tavakol \cite{CET} (CET) and Hosoya and Buchert (HB) \cite{HB} (see comprehensive study of these two proposals in \cite{susslar} and other related research in \cite{sussAN,part1,bolstoeg,misra,Marozzi,Li}).    

In a cosmological context the FLRW models emerge in the various proposals of this entropy as a global ``equilibrium state'' (zero entropy growth $\dot\sgr=0$ for all fundamental observers), thus characterising the deviation from homogeneity and isotropy (inherent in structure formation and cosmic expansion) as irreversible processes whose viability can be verified by means of a theoretically consistent test: demanding that astrophysical and cosmological solutions of Einstein's equations (whether exact, numerical, perturbative or post--Newtonian) fulfill the conditions for gravitational entropy growth $\dot\sgr\geq 0$.  

Considering the empirical success of the $\Lambda$CDM paradigm \cite{paradigm1,paradigm2}, as well as the fact that numerical simulations \cite{nbody} reveal a CDM density distribution made of a web of local cosmic voids of $\sim 50-100$ Mpc size, we believe that these void structures within a concordance cosmology background provide an ideal non--perturbative framework to probe the evolution of the CET entropy proposal, thus complementing a recent study by Marozzi {\it et al} \cite{Marozzi} and earlier work \cite{CET,Li} on perturbative studies appropriate for large scale structure. A convenient non--perturbative approach follows by considering idealised spherically symmetric CDM voids described by Lema\^\i tre--Tolman--Bondi (LTB) dust models with negative spatial curvature \cite{kras,book}. In particular, we aim at looking at the effect of a nonzero cosmological constant by comparing the CET entropy growth predictions for voids whose asymptotic FLRW background is a $\Lambda$CDM model (hence $\Lambda>0$) and an FLRW background model with $\Lambda=0$ ({\it i.e.} spatially flat Einstein de Sitter or open FLRW with negative spatial curvature).  While voids in an FLRW background with $\Lambda=0$ can be ruled out on the grounds of compliance with observations, it is nevertheless important from a theoretical point of view to examine the suppression effect of negative spatial curvature on the growth of the gravitational entropy.

We have shown in previous work \cite{susslar,sussAN} that, as long as the decaying mode is subdominant, a growing CET entropy $\dot\sgr\geq 0$ is a generic feature of LTB models, whether ever expanding or collapsing and irrespective of the void or over--density profiles or the existence of an FLRW background. We also showed that for ever expanding models the rate of growth $\dot\sgr$ is eventually suppressed for all fundamental observers, with a decay $\dot\sgr \to 0$ in the asymptotic time range. We extend this work by computing, through analytic asymptotic expressions, the specific functional form of the entropy growth suppression. The result is a rapid exponential decay of  $\dot\sgr$ for voids in a $\Lambda$CDM background and a slow logarithmic decay for voids in FLRW backgrounds with $\Lambda=0$ (Einstein de Sitter or open FLRW). As a consequence, the CET entropy itself, $\sgr$, which follows from a suitable integral of the time rate $\dot\sgr$ along the worldlines of fundamental observers, only reaches a finite (position dependent) asymptotic terminal ``saturation''  value for voids in a $\Lambda$CDM background, diverging logarithmically for voids in FLRW backgrounds with $\Lambda=0$.   

In order to complement the analytic work described above and to obtain quantitative results, we  examine (numerically) the evolution of the CET entropy $\sgr$ and its rate of growth $\dot\sgr$ for 50--100 Mpc size voids with the three FLRW backgrounds ($\Lambda$CDM, open FLRW and Einstein de Sitter), all evolving from the same plausible generic linear initial conditions (at the last scattering surface) and with a background value for $H_0$ compatible with cosmic age constraints. We obtain results consistent with the asymptotic analytic expressions. We find that the rate of growth $\dot\sgr$ is determined by the amplitude of initial fluctuations of the negative spatial curvature and is basically insensitive to initial fluctuations of the CDM density. The growth suppression proceeds very slowly in the void in an EdS background, so $\sgr$ is still growing at a fast rate in our cosmic time $t_0$. For the voids in the $\Lambda$CDM and open FLRW backgrounds a noticeable suppression of $\dot\sgr$ is already present at relatively early times $t\sim t_0/2$, proceeding at a faster rate for the $\Lambda$CDM background, though $\sgr$ itself is still larger at $t=t_0$ than in the void with an open FLRW background. However, for cosmic times beyond $t_0$ the growth suppression becomes substantially larger in the void with a $\Lambda$CDM background, with $\sgr$ reaching a terminal value at $\sim 2t_0$, while $\sgr$ keeps growing at a very slow logarithmic rate in the void with an open FLRW background. Hence, negative spatial curvature does suppress the CET entropy growth, but not at a sufficiently fast rate to allow for the asymptotic convergence of $\sgr$.

The analytic and numeric results summarised above are qualitatively analogous to and consistent with those obtained in the linear perturbative analysis of \cite{Marozzi}, who find that  the CET entropy growth is suppressed ({\it i.e.} ``saturated'') only for dust perturbations in a $\Lambda$CDM background, with this entropy diverging asymptotically for their ``CDM Universe'' (an Einstein de Sitter background), as these authors did not examine the effects of negative spatial curvature that would arise from an open FLRW background with $\Lambda=0$. Since the exact fluctuations that we used to examine the non--perturbative evolution of $\sgr$ and $\dot\sgr$ reduce to standard cosmological perturbations in the linear regime \cite{SHDG}, we examine the entropy growth in a linear regime that is fully consistent with spherical dust perturbations in the isochronous comoving gauge. As expected, the rate of growth in this regime is much smaller than in the non--linear regime.          

The CET ``gravitational temperature'' associated with the terminal equilibrium value of the CET entropy is in all cases proportional to the Hubble expansion scalar, hence we can identify for local voids in a $\Lambda$CDM background an asymptotic equilibrium temperature proportional to $\sqrt{\Lambda}$, whereas for the open FLRW background this terminal temperature is zero.

The paper is organised as follows. In section \ref{LTBstuff} we briefly present LTB models described in terms of covariant quasi-local scalars (the ``q--scalars''). We introduce in section \ref{CETstuff} the CET entropy and its associated gravitational temperature. Initial conditions at the last scattering time and dimensionless variables are introduced in section \ref{DimVars}. An analytic study of the rate of growth of the CET entropy and temperature in the asymptotic time regime is provided in section \ref{Asymptotics}, while in section \ref{Suppression} we examine the suppression of the CET entropy integral and its asymptotic time limits. In section \ref{Numerics} we examine through  numerical examples of void structures the evolution of the CET entropy and its rate of growth, with a detailed comparison with a linear perturbative approach in section \ref{Comparison}. Finally, we summarise in section \ref{Conclusion} our results and conclusions.

\section{LTB dust models.}\label{LTBstuff}

A convenient parametrization for the LTB dust models is furnished by the following FLRW--like metric (we use units with $G=c=1$):  
\ba 
\fl \dd s^2 =-\dd t^2+ a^2\left[\frac{\Gamma^2}{1-\KKqls r^2}\dd
  r^2+r^2\left(\dd\vartheta^2+\sin^2\vartheta\dd\varphi^2\right)\right],
\label{ltb2}\\
\fl \Gamma = \Gamma(t,r)=1+\frac{r\,a'}{a},\qquad a'=\frac{\partial a}{\partial r},\label{Gdef}\\
\fl a=a(t,r) \quad \hbox{satisfies}\quad \dot a^2 = \frac{8\pi}{3}\left(\frac{\rhoqls}{a}+\Lambda\,a^2\right)-\KKqls,\qquad \dot a=\frac{\partial a}{\partial t},\label{aGdef}
\ea 
where the $r$--dependent functions $\KKqls=\KK_q(\tls,r) $ and $\rhoqls =\rho_q(\tls,r)$ are defined further ahead in (\ref{Aqdef}). The subindex ${}_{\textrm{\tiny{LS}}}$ will denote henceforth evaluation at the last scattering time $t=\tls\approx 4\times 10^5$ ys. The radial coordinate has been chosen so that $\als=\Gals=1$ hold for all $r$.

The standard procedure in most applications of LTB models is to use the solutions (analytic or numerical) of the Friedman equation (\ref{aGdef}) (in various parametrisations) to determine $a$ and $\Gamma$ and then compute all relevant quantities (usually the present cosmic time $t=t_0$ is used as reference fiducial time). An alternative approach is to consider the covariant fluid flow scalars of the models: the energy density $\rho$, the Hubble scalar $\HH\equiv \Theta/3$ with $\Theta=h_a^b\nabla_b u^a$ and the spatial curvature $\KK=\RR/6$ with $\RR$ the Ricci scalar of hypersurfaces of constant $t$ (orthogonal to $u^a$). However, the ``quasi--local'' scalars (to be denoted by ``q--scalars'') and their fluctuations  provide a set of equivalent and useful covariant scalars that are specially suited to probe the CET entropy \cite{susslar} and reduce in a linear regime to gauge invariant spherical dust perturbations in the comoving isochronous gauge \cite{SHDG}. These scalars (which will be denoted by the subindex ${}_q$) are defined as
\footnote{The q--scalars $A_q$ are closely related to weighted proper volume averages of the scalars $A$. They have been very useful to study theoretical properties of LTB models \cite{RadAs,RadProfs,part1,part2,sussmodes} and Szekeres models \cite{sussbol}. Their properties are comprehensively examined in these references. } 
\begin{equation}\fl A_q =\frac{\int_0^r{A R^2 R' \dd\bar r}}{\int_0^r{R^2 R' \dd\bar r}}\qquad \Rightarrow\qquad \rho_q = \frac{\rhoqls}{a^3},\qquad \HH_q=\frac{\dot a}{a},\qquad \KK_q=\frac{\KKqls}{a^2},\label{Aqdef}\end{equation}
where $R=a r$ is the area distance, and they are related to the ``standard'' covariant scalars $A=\rho,\,\HH,\,\KK$ through their fluctuations 
\begin{equation}\DDa = A-A_q=\frac{rA'_q}{3\Gamma}=\frac{1}{R^3}\int_0^r{A'\,R^3\,\\\bar r}\dd\bar r.\label{Dadef}\end{equation}
The q--scalars and their fluctuations satisfy the following algebraic constraints
\begin{equation}  \HH_q^2 = \frac{8\pi}{3}(\rho_q+\Lambda) -\KK_q,\qquad
2\HH_q\DDH = \frac{8\pi}{3}\DDrho - \DDKK.\label{constr}\end{equation}
The shear and electric Weyl tensors are nicely expressed in terms of the fluctuations of the density and Hubble scalar:
\begin{equation}\sigma_a^b=-\DDH\,\eigf_a^b=\frac{\dot \Gamma}{3\Gamma}\,\eigf_a^b,\qquad E_a^b=-\frac{4\pi}{3}\DDrho \eigf_a^b=\Psi_2\,\eigf_a^b,\label{sigE}\end{equation}
where $\Psi_2$ is the conformal Newman--Penrose invariant and $\eigf_a^b=h_a^b-3n_a n^b=\hbox{diag}[0,-2,1,1]$ (with $n_a$ a unit vector orthogonal to $u^a$ and to the orbits of SO(3)). As shown in previous literature \cite{part1,RadAs,RadProfs,part2,sussmodes,sussbol,sussDS2}, the q--scalars and their exact fluctuations provide a representation of covariant scalars that fully determines the dynamics of LTB models (either analyticaly or numerically). 
 
\section{The CET gravitational entropy in LTB models.}\label{CETstuff}   

The CET gravitational entropy is defined from an ``effective'' energy momentum tensor ${\cal T}^{ab}$ for a suitable geometric field associated with the Weyl tensor (the ``free'' gravitational field 
\footnote{CET obtain the second order ``effective'' energy--momentum tensor ${\cal T}^{ab}$ through an irreducible algebraic decomposition of the Bell--Robinson tensor, the only fully symmetric divergence--free tensor that can be constructed from the Weyl tensor. See comprehensive discussion in \cite{CET}.}).
For LTB models (which are Petrov type D spacetimes or ``Coulomb--like'' fields), this tensor takes the form \cite{CET}:
\ba \frac{{\cal T}^{ab}}{8\pi} = \rhogr u^au^b + \pgr h^{ab}+2q^{(a}_{\tiny{\textrm{gr}}} u^{b)}+\Pigr,\label{Tabgr}\ea
where $u^a$ is the 4--velocity of the matter source and the  ``gravitational'' density, pressure, viscosity and heat flux $\rhogr,\,\pgr,\,\Pigr,\,\qgr$ are given in terms of the conformal invariant $\Psi_2$ (directly related to the density fluctuation $\DDrho$ by (\ref{sigE})) as
\ba \fl 8\pi\rhogr = \frac{8\pi}{3}\alpha\,|\DDrho|,\qquad  8\pi\Pigr = \frac{4\pi}{3}\alpha\,|\DDrho|\,\PP^{ab},\qquad \pgr=\qgr=0,\label{effective}
\ea
where $\PP^{ab}=x^ax^b+y^ay^b-n^a n^b+u^au^b$ for the canonical orthonormal tetrad $\{u^a,n^a,x^a,y^a\}$ associated with (\ref{ltb2}) and $\alpha$ is a constant to get the right units. By setting up a formal analogy between ${\cal T}^{ab}$ in (\ref{Tabgr}) and the energy--momentum tensor of a ``reference'' dissipative hydrodynamical source with 4--velocity $u^a$ \cite{rund}, CET obtain a Gibbs equation analogue for the gravitational entropy growth, which for LTB models takes the from (see details in \cite{CET} and \cite{susslar}):
\ba
\fl \Tgr\dotsgr &=&  (\rhogr V)\dot{}= -V\sigma_{ab}\left[\Pigr+\frac{\rho}{\alpha\,|\DDrho|}E^{ab}\right]= -\frac{2\pi\alpha}{3}\,\frac{\partial}{\partial t}\left(|\DDrho|\,a^3\,\Gamma\right) \nonumber\\
\fl  &=& -2\pi\alpha\rho_q\,a^3 \Gamma\DDH\,\frac{\DDrho}{|\DDrho|},
\label{CETc12}
\ea
where we used (\ref{Aqdef}), (\ref{FFq1})--(\ref{FFq4}), (\ref{FFq5}), the tensors  $\sigma_{ab},\,E^{ab}$ follow from (\ref{sigE}), $V=\ell^3=a^3\Gamma$ is the local volume defined by $\dot\ell/\ell=\HH$ and the ``gravitational'' temperature $\Tgr$ was given by CET in \cite{CET} as
\begin{equation} \Tgr = \frac{\left|\HH+\sigma_{ab}n^an^b\right|}{2\pi} = \frac{|\HH_{q}+3\DDH|}{2\pi},\label{Tgr}  \end{equation}
on the grounds that it reduces to the semi--classical Unruh and Hawking temperatures in the appropriate limits (see discussion in \cite{CET}).  The terms inside the brackets in the right hand side of (\ref{CETc12}) describe the  ``effective'' dissipation produced by the CET entropy through the geometric fluid associated with (\ref{effective}) (notice that there is no exchange of energy or momentum between this formal fluid and the LTB dust source). Since we have $\sigma_{ab}=0$ for FLRW models, the latter define a global ``gravitational'' equilibrium state characterized by $\Tgr=|\HH|/(2\pi)>0$ and $\dotsgr=0$ for all $t$ and all fundamental observers.   

Assuming that $\Gamma>0$ holds to avoid shell crossing singularities \cite{RadProfs,sussmodes}, we obtain from (\ref{CETc12}) the  necessary and sufficient condition for entropy growth along the worldlines of fundamental observers \cite{susslar}:
\begin{equation} \dotsgr\geq 0 \quad \Leftrightarrow \quad \DDrho\DDH\leq 0.\label{CETcond1}\end{equation}
which takes the form of a negative correlation between fluctuations of the energy density and the Hubble scalar with respect to their q--scalars. 

As argued in \cite{susslar}, the CET entropy growth condition (\ref{CETc12}) is only the time component ({\it i.e.} component projected to the 4--velocity) of the one--form
\begin{equation}\Tgr\,\bdd\sgr= \frac{2}{3}\pi\alpha \,\bdd(\rhogr V).\label{sgrform}\end{equation}
whose integrability conditions for LTB models place the following restriction on the radial dependence of $\sgr$ (see details in \cite{susslar}):
\begin{equation} \sgr' = \frac{(\Tgr^2)'}{(\Tgr^2)\dot{}}\, \dot\sgr= \frac{\HH'_q +3(\DDH)'}{\dot\HH_q+3\dot\DDH} \, \dot\sgr,\label{sgrprime}\end{equation}
which must be considered when looking at the integration of $\bdd \sgr$ in (\ref{sgrform}) (see section \ref{Suppression}). 

\section{Initial conditions and dimensionless variables.}\label{DimVars}     

Since we will consider initial conditions at the last scattering time $t=\tls$, it is useful to  normalise all initial value functions with respect to the Hubble length scale $\Hls$ of the FLRW background at this time (an over--bar will denote henceforth  FLRW background quantities). In particular, the initial CDM density and spatial curvature (standard scalars not q--scalars) and cosmological constant can be given by   
\bse\ba\fl  2\muls=\frac{8\pi\rhols}{3\Hls^2}=\Ommls+m_1\epsilon^{(m)}(r),\quad \kls=\frac{\KKls}{\Hls^2}=\Omkls+k_1\epsilon^{(k)}(r),\label{initconds1a}\\
\fl \lambda = \OmLls=\OmLO\frac{\bar H_0^2}{\Hls^2}=1.37\times 10^{-9} \quad (\OmLO=0.73,\quad \bar H_0c=69\hbox{km/s Mpc}),\label{initconds1b}
\ea\ese 
where $\Hls,\,\Ommls,\,\OmLls$ and $\Omkls=\Ommls+\OmLls-1$ characterise the FLRW background at $\tls$, the smooth functions $\epsilon^{(m)}(r),\,\epsilon^{(k)}(r)$ are selected to fulfil regularity conditions (see \cite{part2}) and to satisfy $\epsilon^{(m)}(0)=\epsilon^{(k)}(0)=1$ and $\epsilon^{(m)},\,\epsilon^{(k)}\to 0$ as $r\to \infty$, while the constants $m_1,\,k_1$ comply with $(|m_1|,\,|k_1|)\ll 1$. The  q--scalars and fluctuations associated to $\muls$ and $\kls$ are
\bse\ba \fl \muqls =\frac{4\pi\rhoqls}{3\Hls^2}=\frac{\Ommls}{2}+m_1\epsilon^{(m)}_q,\quad \kqls =\frac{8\pi\KKqls}{3\Hls^2}=\Omkls+k_1\epsilon^{(k)}_q,\label{initconds2a}\\
\fl \hqls =\frac{\HHqls}{\Hls}=\left[2\muqls-\kqls+\lambda\right]^{1/2}=\left[1+2m_1\epsilon_q^{(m)}-k_1\epsilon_q^{(k)}\right]^{1/2},\label{initconds2b}\\
\fl \hqls \approx 1+m_1\epsilon_q^{(m)}-\frac{k_1}{2}\epsilon_q^{(k)},\qquad \OmqLls = \frac{\OmLls}{\hqls^2}\approx \OmLls, \label{initconds2c}\\
\fl \DDmuls = \frac{r}{3}\muqls',\qquad \DDkls =\frac{r}{3}\kqls',\qquad \DDhls =\frac{r}{3}\hqls'=\frac{2\DDmuls-\DDkls}{2\hqls},\label{initconds2d}\ea\ese
where we applied (\ref{Aqdef})--(\ref{constr}) to (\ref{initconds1a})--(\ref{initconds1b}) and $\epsilon^{(m)}_q,\,\epsilon^{(k)}_q$ are obtained from $\epsilon^{(m)},\,\epsilon^{(k)}$ from the integral (\ref{Aqdef}) evaluated at $\tls$ ({\it i.e.} with $R=r$, since $\als=\Gals=1$).

For further studying the evolution of the CET entropy for the initial conditions given above, we introduce the dimensionless  time and radial coordinate normalised with respect to the Hubble scale
\begin{equation} \tau = \Hls ct-\frac{2}{3}=\frac{2}{3}\left(\frac{t}{\tls}-1\right),\qquad \chi = \Hls\,r=\frac{2 r}{3c\tls},\label{tau}\end{equation}
so that the initial time slice $t=\tls$ is given by $\tau=0$ and present day cosmic time $t_0\sim 13.7$ Gys is $\tau_0\approx 2.2\times 10^4$, while $\chi=1$ corresponds to the Hubble radius $r=1/\Hls=3 c\tls/2\approx 0.19$ Mpc. 

Regarding the entropy growth, we recall that the term $2\pi\alpha$ in (\ref{CETc12}) has units of entropy times length (after writing $2\pi\rho_q$ as $2\pi G\rho_q/c^2$). Hence, considering that $\rho_q a^3=\rhoqls$, the components of the CET entropy one--form (\ref{sgrform}) in (\ref{CETc12}) and (\ref{sgrprime}) can be written in terms of dimensionless quantities as 
\ba\fl \frac{\partial\tilde\sgr}{\partial\tau}=-\frac{3}{2}\,\muqls\,\frac{\Gamma\,\DDh\,\DDmu}{|h_q+3\DDh|\,|\DDmu|},\label{sgrt2}\\
\fl \frac{\partial \tilde\sgr}{\partial\chi}=-\frac{3}{2}\,\muqls\,\frac{\left[h_q+3\DDh\right]_{,\chi}}{\left[h_q+3\DDh\right]_{,\tau}}\times \frac{\Gamma\,\DDh\,\DDmu}{|h_q+3\DDh|\,|\DDmu|},\label{sgrr2}\ea
where $\tilde\sgr\equiv \sgr/\sigma$, with $\sigma$ being a suitable quantity with entropy units (its theoretical role is discussed in section \ref{Conclusion}) and 
\begin{equation}\fl h_q=\frac{\HH_q}{\Hls},\qquad  \DDh=h-h_q=\frac{\DDH}{\Hls},\qquad \DDmu=\mu-\mu_q=\frac{4\pi \DDrho}{3\Hls^2},\label{dimvars}\end{equation}
generalise for all $\tau$ the initial value functions $\hqls,\,\DDhls$ and $\DDmuls$.  The two components of $\bdd\sgr$ in (\ref{sgrt2}) and (\ref{sgrr2}) must be considered separately when looking at $\sgr$ along comoving observers ($\partial\tilde\sgr/\partial\tau$ for $\chi$ constant) and along radial rays of hypersurfaces orthogonal to the 4--velocity ($\partial\tilde\sgr/\partial\chi$ for $\tau$ constant).

\section{Asymptotic entropy growth and temperature.}\label{Asymptotics}    

We proved in \cite{susslar} that $\partial\tilde\sgr/\partial\tau>0$ holds in the time asymptotic regime of fundamental observers in generic LTB models with negative spatial curvature in the case $\Lambda=0$. We review this result here and prove that it also holds for the case $\Lambda>0$. We also examine the asymptotic evolution of the gravitational temperature $\Tgr$. For this purpose we need the time asymptotic forms of the metric functions $a$ and $\Gamma$, which can be obtained from the asymptotic range $a\gg 1$ (or $t/\tls \gg 1$) of the quadrature of the Friedman equation (\ref{Gdef}) 
\begin{equation}\fl \hqls\tau = F(a,\Omqmls,\OmqLls)=\int_1^a{\frac{\sqrt{\xi} d\xi}{\left[\Omqmls-\Omqkls\,\xi+\OmqLls\,\xi^3\right]^{1/2}}},\label{quadrature}\end{equation}
where the Omega factors are 
\begin{equation}\fl\Omqmls = \frac{2\muqls}{\hqls^2},\qquad \OmqLls=\frac{\lambda}{\hqls^2},\qquad \Omqkls=\frac{\kqls}{\hqls^2}=\Omqmls+\OmqLls-1,\end{equation}
and the lower bound of the integral corresponds to $t=\tls$ (or $\tau=0,\,a=\als=1$). For ever expanding dust layers (see conditions for this when $\Lambda>0$ in \cite{sussDS2}) the late time forms for the metric functions $a$ and $\Gamma$ are
\ba \fl a\approx \exp\left(\sqrt{\lambda}\,\hqls\,\tau\right),\quad \Gamma \approx 1+3\DHls-\frac{3 [(\Omqmls+\OmqLls)\Dhls-\frac{1}{2}\Omqmls\,\Dmuls]}{a^2},\quad (\Lambda>0),\nonumber\\
\fl a\approx \sqrt{1-\Omqmls}\,\hqls\,\tau,\qquad \Gamma\approx 1+\frac{3}{2}\Dkls+\frac{3\Omqmls\,\left[\Dmuls-\frac{3}{2}\Dkls\right]}{1-\Omqmls}\frac{\ln\,a}{a},\qquad (\Lambda =0),\nonumber\\\fl \label{aGas}
\ea
where we have introduced the relative initial fluctuations 
\begin{equation}\fl  \Dmuls =\frac{\DDmuls}{\muqls},\qquad \Dkls =\frac{\DDkls}{\kqls},\qquad \Dhls =\frac{\DDhls}{\hqls}=\frac{\Omqmls}{2}\Dmuls-\frac{\Omqkls}{2}\Dkls.\label{deltadef}\end{equation}
Inserting the forms (\ref{aGas}) into (\ref{Aqdef}) and (\ref{Dadef}) (expressed in terms of dimensionless quantities) and considering only leading  terms in $1/a$ leads after a long algebraic manipulation to the time asymptotic form of the product of fluctuations
\bse\ba\fl  \DDmu\,\DDh &\approx& -\frac{\hqls^3\Omqmls}{4\sqrt{\OmqLls}}\,\frac{\left[\Omqmls\Dmuls-2(\Omqmls+\OmqLls)\Dhls\right]\left[\Dmuls-3\Dhls\right]}{(1+3\Dhls)^2\,a^5}\qquad\qquad\nonumber\\
\fl &\approx& -\frac{(\Dmuls-2\Dhls)(\Dmuls-3\Dhls)}{\sqrt{\OmqLls}\,a^5},\qquad\qquad\qquad\qquad (\Lambda>0),\label{DmDh1}\\
\fl \DDmu\,\DDh &\approx&  -\frac{\hqls^3(\Omqmls)^2\left[(2+\Omqmls)\Dmuls-6\Dhls\right]^2}{4\sqrt{1-\Omqmls}\left[\Omqmls\left(1+\frac{3}{2}\Dmuls\right)-1-3\Dhls\right]^2}\,\frac{\ln a}{a} \nonumber\\
\fl &\approx& -\frac{\left(\Dmuls-3\Dhls\right)^2}{\sqrt{1-\Omqmls}\,\left(\Dmuls-\frac{3}{2}\Dhls\right)^2}\,\frac{\ln a}{a},\qquad\qquad\qquad (\Lambda=0),\label{DmDh2}\ea\ese 
where we used the approximations $\Omqmls\approx 1,\,\hqls\approx 1,\,\OmqLls\ll 1$.  The expansions (\ref{DmDh1})--(\ref{DmDh2}) determine the late time sign of $\partial\tilde\sgr/\partial\tau$ through condition (\ref{CETcond1}) for arbitrary radial profiles of initial value functions specified at $\tls$.

Considering the fact that $\DDh\sim a^{-2}\to 0$ holds as $\tau\to\infty$ \cite{susslar,sussmodes}, the gravitational temperature associated to the asymptotic equilibrium state defined by $\partial\tilde\sgr/\partial\tau\to 0$ is given by $\Tgr \to h_{\textrm{\tiny{inf}}}/2\pi$ where $h_{\textrm{\tiny{inf}}}$ is the limit of $h=\HH/\Hls$ as $\tau\to\infty$. We look at the cases $\Lambda=0$ and $\Lambda>0$ separately below.

\subsubsection{The case $\Lambda=0$.} As proven in \cite{susslar}, equations (\ref{CETcond1}) and (\ref{DmDh2}) confirm that the fulfilment of $\partial\tilde\sgr/\partial\tau>0$ in the asymptotic time range of fundamental observers is a generic feature in hyperbolic models with negative spatial curvature ($0<\Omqmls<1$ implies  $\kappa_q<0$ for all $t$), irrespective of the nature of the density profile at  $t=\tls$ or of the existence of a decaying mode which is already subdominant at $\tls$ and is completely negligible for late times \cite{susslar,sussmodes}. We have $h_{\textrm{\tiny{inf}}}=0$,  hence $\Tgr \to 0$, and thus the asymptotic equilibrium state is characterised by zero gravitational temperature.   
     
\subsubsection{The case $\Lambda>0$.} Considering a $\Lambda$CDM background with $\Omkls=0$, together with the relation between the signs of the radial gradients and the fluctuations in (\ref{Dadef}), and assuming monotonic (but otherwise unrestricted) radial density and spatial curvature profiles at the initial slice $\tls$, we have:  
\begin{itemize}
\item Initial void profile: $\muqls'\geq 0$, hence $\Dmuls\geq 0$ (density minimum at the void centre), but $\hqls'\leq 0$ and thus  $\Dhls\leq 0$ (maximal positive Hubble expansion at the centre of the void). 
\item Initial over--density profile: $\muqls'\leq 0$, hence $\Dmuls\leq 0$ (density maximum at the clump centre), but $\hqls'\geq 0$, hence $\Dhls\geq 0$ (minimal positive Hubble expansion at the clump centre).
\end{itemize}
and thus it is straightforward to show from (\ref{CETcond1}) and (\ref{DmDh2}) that $\partial\tilde\sgr/\partial\tau>0$ holds in the asymptotic time range. While the numerical examples in \cite{sussDS2} show this result, we prove it here analytically.  Since we have
\begin{equation} 2\pi \Tgr\to \sqrt{\OmqLls}\hqls=\sqrt{\frac{8\pi}{3}\Lambda},\end{equation}
the CET entropy provides (irrespective of initial conditions) a nice interpretation of $\Lambda$ as proportional to the asymptotic gravitational temperature of the terminal equilibrium state.   
     
\section{The suppression and ``saturation'' of the CET entropy.}\label{Suppression}

The correlations in (\ref{DmDh1}) and (\ref{DmDh2}) suggest a dramatic asymptotic suppression of the CET entropy growth for fundamental observers when $\Lambda >0$. We still need to verify this suppression (and the asymptotic suppression in the radial direction) from the general expressions (\ref{sgrt2}) and (\ref{sgrr2}). This information will allow us to infer the general asymptotic behaviour (for fundamental observers) of $\sgr$ itself.   

\subsection{Entropy suppression for fundamental observers.}\label{Suppression1}

Considering the full form of $\partial\tilde\sgr/\partial\tau$ from (\ref{sgrt2}), using the expansions (\ref{aGas}) and the products of $\DDrho\DDH$, the asymptotic form of the rate of decay of entropy growth for fundamental observers ($\chi$ constant) is given by:
\ba \fl \frac{\partial\tilde\sgr}{\partial\tau} \approx \frac{3}{8} \frac{\Omqmls\left[\left(\Omqmls+\OmqLls\right)\DDkls-\Omqkls\DDmuls\right]}{\OmqLls\,a^2}\approx \frac{3\DDkls}{8\OmqLls}\hbox{e}^{-2\sqrt{\OmLls}\tau},\quad (\Lambda>0),\nonumber\\
\fl \label{sgras1}\\
\fl \frac{\partial\tilde\sgr}{\partial\tau} \approx \frac{3}{4} \frac{\Omqmls \left[3\Omqmls \DDkls -4\Omqkls\DDmuls \right]}{(\Omqkls)^2}\,\frac{\ln a}{a}\approx \frac{9\DDkls}{4(\Omqkls)^2}\,\frac{\ln \tau}{\tau},\quad (\Lambda=0),\nonumber\\\fl \label{sgras2}\ea
which reveals a decay to zero of entropy growth rate for fundamental observers located in voids in all FLRW backgrounds (irrespective of whether $\Lambda>0$ or $\Lambda=0$). However, this decay occurs at a fast exponential rate in voids with a $\Lambda$CDM background, while it is a slow logarithmic decay $\ln t/t$ for voids in any FLRW background with $\Lambda=0$. In order to explore the  consequences of these decay rates, we introduce the entropy integral for fundamental observers ({\it i.e.} the line integral of the one--form $\bdd s$ projected along the integral curves of the 4--velocity) 
\begin{equation}  \tilde\sgr(\tau)|_\chi\equiv \int_{{\bf{u}}}{(\bdd \tilde\sgr\cdot {\bf u})\,\dd\tau}=\phi(\chi)+\int_{0}^\tau{\left[\frac{\partial\tilde\sgr}{\partial\tau}\right]_{\chi\hbox{\tiny{=const}}} \dd \tau},\label{sgrFO}\end{equation}
where $\phi(\chi)$ is an ``integration constant'' that will be related further ahead to the initial entropy state. 

Evidently, the entropy integral (\ref{sgrFO}) only converges asymptotically to a finite terminal equilibrium or ``saturation'' value for all fundamental observers in voids whose FLRW background is $\Lambda$CDM (or any FLRW model with $\Lambda>0$). For fundamental observers in voids whose background has $\Lambda=0$ the logarithmic decay in (\ref{sgras2}) leads to a diverging asymptotic limit
\footnote{In \cite{susslar} we concluded, mistakenly, that the time integral of $\sgr$ converges asymptotically for hyperbolic LTB models with $\Lambda=0$ because $\dot\sgr\to 0$ holds as $t\to\infty$. The fact that we now show that $\sgr$ diverges logarithmically for voids in an open FLRW background corrects this mistake, as these voids are  based on such hyperbolic models.}
\begin{equation}\tilde\sgr(\tau)|_\chi\sim \int{(\ln\,\tau/\tau) dt}\sim [\ln\,\tau]^2\to \infty\quad \hbox{as}\quad \tau\to\infty.\end{equation}
These results coincide qualitatively with the perturbative analysis of \cite{Marozzi} (though an open FLRW background was not considered in this reference).

\subsection{Entropy suppression in the radial direction.}\label{Suppression2}

Considering the radial component of $\bdd\sgr$ given by (\ref{sgrprime}) and (\ref{sgrr2}), it was shown in section 9 of \cite{susslar} that $[\partial\sgr/\partial\chi]_\tau\to 0$ as $\chi\to\infty$ holds for all LTB models with $\Lambda=0$ along radial rays: {\it i.e.} curves with $(\tau,\vartheta,\varphi)$ constant, which are  
spacelike geodesics of the LTB metric and integral curves of the spacelike unit vector $n^a =\sqrt{g^{rr}}\delta^a_r$ orthogonal to $u^a$ and to the orbits of SO(3).   

Following \cite{RadAs}, it is straightforward to extend the above mentioned proof to the case of LTB models with $\Lambda>0$ admitting a FLRW background in the asymptotic radial direction, as the metric function $a$, the covariant scalars $\rho,\,\KK,\,\HH$ and their associated q-scalars tend as $\chi\to\infty$ to their background values $\bar a,\,\bar\rho,\,\bar H,\,\bar\KK$ (an over--bar denotes FLRW quantities), while the fluctuations $\DDa$ vanish in this limit. Considering the polynomial asymptotic forms in equations (84)--(86) and (88) of \cite{susslar} together with (\ref{Aqdef})--(\ref{constr}) we have in the asymptotic range of radial rays in arbitrary slices of constant $\tau$  
\begin{equation} \HH_q\sim \bar H +\bar H_1\,\chi^{-\nu},\qquad \DHH=\frac{\DDH}{\HH_q}\sim -\frac{\nu\bar H_1}{3\bar H}\chi^{-\nu}\end{equation}
where $\nu>1$ is a constant, $\bar H=\bar H(\tau)=\dot{\bar a}/\bar a$ is the background Hubble scalar and $\bar H_1=\bar H_1(\tau)$, leading to
\begin{equation}\fl \left[\frac{\partial \sgr}{\partial\chi}\right]_{\tau=\hbox{\tiny{const.}}} \sim 3\bar H\DHH\left[\frac{\DHH}{\chi}+\frac{\partial}{\partial\chi}\DHH\right]\approx -\frac{\nu(\nu-1)\,\bar H_1^2}{\bar H}\,\chi^{-2\nu}.\end{equation}     
Proceeding along the lines of (\ref{sgrFO}), we define the radial entropy integral as the line integral of the one--form $\bdd \tilde\sgr$ along the radial rays:
\begin{equation}  \tilde\sgr(\chi)|_\tau\equiv \int_{{\bf{n}}}{(\bdd \tilde\sgr\cdot {\bf n})\,\dd\chi}= \int_{0}^\chi{\left[\frac{\partial\tilde\sgr}{\partial\chi}\right]_{\tau=\hbox{\tiny{const}}} \dd \chi}
,\label{sgrRR}\end{equation}
where $[\partial\tilde\sgr/\partial\chi]_{\tau=\hbox{\tiny{const}}}$ is given by ({\ref{sgrr2}) and $\tau\geq 0$ is now an arbitrary fixed finite parameter and we set the ``integration constant'' $\sgr(\tau,0)=\sgr(\tau)|_{\chi=0}$ to zero (because the fluctuations $\DDmu,\,\DDh$ vanish at the symmetry centre).  Evidently, for the polynomial decay considered in \cite{susslar} the integral $\tilde\sgr(\chi)|_\tau$ above converges to a finite asymptotic value value as $\chi\to\infty$ for radial rays in all FLRW backgrounds. This is a generic behaviour for a wide range of initial conditions and would only change if the latter are chosen such that $a$ and the covariant scalars decay logarithmically along radial rays (which merely places the restriction to avoid such initial conditions).   
   
\subsection{The Entropy line integral.}\label{Suppression3}

The initial state of the CET entropy  (\ref{sgrFO}) for fundamental observers is the CET entropy evaluated at $\tls$ for all $\chi$. Hence, we can identify the integration constant in (\ref{sgrFO}) with this initial state given by the radial integral (\ref{sgrRR}) for $\tau=0$ (or $t=\tls$)
\begin{equation} \tilde\sgrls(\chi)= \tilde\sgr(\chi)|_{0}=\int_0^r{\left[\frac{\partial\tilde\sgr}{\partial\chi}\right]_{\tau=0} \dd \chi}.\label{sgrLS}\end{equation}
Therefore, for any given fundamental observer at cosmic time $t=t^*\geq \tls$ ($\tau=\tau^*>0$) comoving along a worldline marked by arbitrary fixed $\chi=\chi^*\geq 0$, the gravitational entropy can be defined as the line integral of the CET entropy one--form $\bdd \sgr$ along two paths: a radial ray at $\tau=0$ running from $\chi=0$ to $\chi=\chi^*$ and the comoving worldline marked by $\chi=\chi^*$ running from $\tau=0$ to $\tau^*$, which is the sum of the two entropy line integrals (\ref{sgrLS}) and (\ref{sgrFO})
\footnote{Our integration of the CET entropy one--form $\bdd \sgr$ is well posed and self--consistent regardless of general integrability considerations, as any one--form can be consistently integrated as a line integral along specific paths, even if it is not closed or exact. For the study of the frame dependent CET entropy on LTB models the  physically meaningful paths are: radial rays at the initial time slice (spacelike geodesics and integral curves of $n^a$) and the comoving worldlines (timelike geodesics and integral curves of $u^a$). Discussing the general integrability of the CET entropy one--form $\bdd \sgr$ is beyond the scope of this article and will be examined elsewhere.} 
\begin{equation} \tilde\sgr(\tau^*,\chi^*) = \tilde\sgrls(\chi^*) + \tilde\sgr(\tau^*)|_{\chi^*}.\label{sgrTOT}\end{equation}  
Evidently, if we are interested in the CET entropy referred to fundamental observers the radial entropy integral (\ref{sgrRR})  only contributes to the evaluation of the total CET entropy (\ref{sgrTOT}) in the specification of a different initial entropy state (\ref{sgrLS}) for each observer. As shown in the previous section, this radial integral generically converges, and thus it  merely adds a finite $r$--dependent initial condition term to the time integral $\sgr(\tau)|_\chi$, which will not affect the  qualitative asymptotic behaviour of the CET entropy for fundamental observers that follows from (\ref{sgras1})--(\ref{sgras2}) and (\ref{sgrFO}). In fact, by assuming fundamental observers in bounded comoving domains that are radially asymptotic to an FLRW background, we can state the following general result:
\begin{itemize}
\item $\tilde\sgr$ converges to a finite value for observers in voids with a $\Lambda$CDM background,
\item $\tilde\sgr$ diverges logarithmically for observers in voids with a $\Lambda=0$ background.
\end{itemize}  
which for a wide range of LTB configurations coincides qualitatively with the results of the perturbative analysis of \cite{Marozzi}. 
        
\section{Numerical analysis.}\label{Numerics} 

The asymptotic results of the previous sections reveal that the growth of the CET entropy for fundamental observers is suppressed for generic models, but provide no information on the cosmic times in which this suppression sets in for interesting dust inhomogeneities. To look at these issues properly we need to obtain more detailed information on $\partial\tilde\sgr/\partial\tau$ and its time integral $\tilde\sgr(\tau)|_\chi$ in their full time evolution from suitable initial data at $\tls$ (or $\tau=0$).

\subsection{Evolution equations.}\label{Numerics1}

In the case $\Lambda=0$ and in some cases with $\Lambda>0$ (see \cite{sussDS2}) the form of $\partial\tilde\sgr/\partial\tau$ in (\ref{sgrt2}) can be found analytically from the solutions of the Friedman equation (\ref{aGdef}) given by the quadrature (\ref{quadrature}), as the latter are expressible as elementary functions that can easily be parametrised by the variables $A_q,\,\DDa$ \cite{part1,RadAs,RadProfs,part2,sussmodes,sussbol,sussDS2}. In the general case $\Lambda>0,\,\kqls\ne 0$, this task is better undertaken by means of numerical or semi--analytic solutions of (\ref{aGdef}) (see for example \cite{valk}). However, it is far more efficient for numerical work to compute all quantities directly (through the $A_q,\,\DDa$) by the numerical solutions of the following system of autonomous ODE's \cite{sussDS2}
\footnote{The evolution equations used in \cite{sussDS2} are equivalent to (\ref{FFq1})--(\ref{FFq4}), but were constructed with the $A_q$ and exact dimensionless relative functuations (called ``perturbations'') $\Da = \DDa/A_q$.}
: 
\bse\ba \fl \frac{\partial\mu_q}{\partial \tau} &=& -3 \mu_q\,h_q,\label{FFq1}\\
\fl \frac{\partial\, h_q}{\partial \tau} &=& -h_q^2-\mu_q+\lambda, \label{FFq2}\\
\fl \frac{\partial\,\DDmu}{\partial \tau} &=& -3(h_q+\DDh)\DDmu-3\mu_q\DDh,\label{FFq3}\\
\fl \frac{\partial\,\DDh}{\partial \tau} &=&  -\DDmu-(2h_q+3\DDh)\,\DDh,\label{FFq4}\\
\fl \frac{\partial\,a}{\partial \tau} &=& a\,h_q, \label{FFq5}\\
\fl \frac{\partial\,\Gamma}{\partial \tau} &=& 3\,\Gamma\,\DDh, , \label{FFq6}\ea\ese
where the dimensionless variables $\{\mu_q,\,h_q,\,\DDmu,\,\DDh\}$ and the dimensionless time $\tau$ are defined in (\ref{tau}) and (\ref{dimvars}). 

\subsection{Initial conditions.} \label{Numerics2} 

The system (\ref{FFq1})--(\ref{FFq6}), which is subjected to the algebraic constraints (\ref{constr}), must be integrated numerically for initial conditions (\ref{initconds1a})--(\ref{initconds1b}) and (\ref{initconds2a})--(\ref{initconds2c}) specified at $\tls$. Since our aim is to study the CET entropy growth and its time integral for 50--100 Mpc voids at present cosmic time $\tau_0$, we need to explore which parameters of these initial conditions determine the entropy growth in such present day structures. For this purpose, we remark from looking at the asymptotic expansions (\ref{sgras1})--(\ref{sgras2}) that the late time rate of CET entropy growth is basically determined by the amplitude ($\sim |k_1|$) of spatial curvature initial fluctuation $\DDkls$ and that this growth is insensitive  to the amplitude ($\sim |m_1|$) of density fluctuation $\DDmuls$. However, this sensitivity to initial spatial curvature fluctuations is different when $\Lambda$ is zero and nonzero:
\begin{itemize}
\item If $\Lambda>0$ a larger amplitude of $\DDkls$ (larger $|k_1|$) yields larger entropy growth,
\item If $\Lambda=0$ larger entropy growth occurs for smaller amplitude of $\DDkls$ (smaller $|k_1|$), or equivalently, for smaller deviation from spatial flatness. 
\end{itemize}
As we show further ahead, these patterns are confirmed by the numerical examples. In particular, (\ref{sgras2}) implies that entropy growth should be much smaller in voids in an open FLRW background (negative spatial curvature) than in the spatially flat Einstein de Sitter background, as curvature fluctuations in the latter are expected to be much larger than in the former. 

Since, as follows from numerical trials, different shapes of the admissible functions $\epsilon^{(m)}$ and $\epsilon^{(k)}$ in (\ref{initconds1a}) do not produce significant qualitative changes in the shape of $[\partial\tilde\sgr/\partial\tau]_{\tau}$, we can explore effectively the effects of initial conditions to obtain entropy growth rates for specific void models by varying $k_1$ for a fixed $m_1$, and using the simplest functional form for $\epsilon^{(m)},\,\epsilon^{(k)}$. In particular, we select the following simple polynomial profiles for these functions: 
\begin{equation}\fl
\muls = \frac{\Ommls}{2}+\frac{m_{1}}{1+(\chi/m_2)^2}\qquad 
\kls = \Omkls +\frac{k_{1}}{1+(\chi/k_2)^{3/2}},\label{initconds3}
\end{equation}
where the dimensionless comoving radius $\chi$ is defined in (\ref{tau}).  Since the value of $10^{-3}$ is an appropriate  upper bound for the amplitude of CDM density fluctuations at $\tls$ when using LTB models to describe voids in structure formation scenarios \cite{book}, we have fixed the maximal value $|m_1|\sim 10^{-3}$, leaving the value of $|k_1|$ as a free parameter in the initial conditions to be varied in order to obtain voids with $\sim -0.8$ density contrast of a typical size of $50-100$ Mpc at $\tau_0$. Numerical trials using the ansatz (\ref{initconds3}) with $|m_1|\sim 10^{-3}$ reveal that the range of values $|k_1|\sim 10^{-3}-10^{-2}$ leads to such voids. Hence,  we chose for setting up specific numerical examples the constant parameters $m_1,\,m_2$ and $k_1,\,k_2$ summarised in Table~\ref{Tab:Tab1} that yield the desired voids in an Einstein-de-Sitter background (EdS), an FLRW background with negative curvature and $\Lambda=0$ (open FLRW), and a $\Lambda$CDM background.  The density contrast, Hubble scalar and mass distribution for voids with these parameters are depicted by Fig.~\ref{Fig:Fig1}.
%
\begin{figure}[h] 
\begin{center}
\includegraphics[scale=0.40]{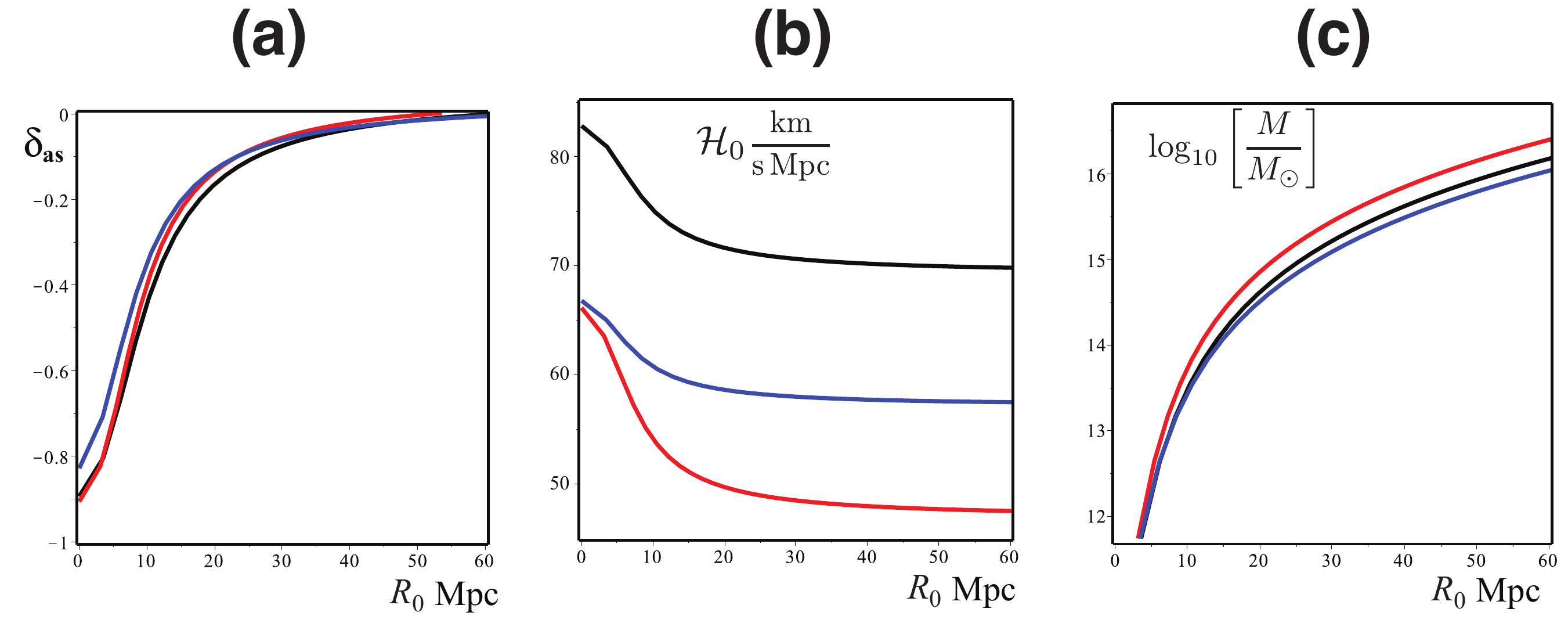}

\end{center}
\caption{{\bf The CDM void examples}. The figure depicts the density contrast (panel (a)), Hubble scalar $\HH_0$ (panel(b)) and mass distribution $M=(4\pi/3)\rho_{q0}R_0^3$ in solar mass units for the void models that follow from the initial conditions listed in Table 1. All curves are evaluated at present cosmic time $\tau_0$ as functions of area distance $R_0=a_0\,r$. The black, red and blue curves (online version) denote the voids in $\Lambda$CDM, EdS and open FLRW backgrounds. Notice that the three void models have practically the same radial distribution of CDM density and mass. However, the Hubble scalar is different for each void, as its background value (see Table 1) has been selected to meet cosmic age constraints.}
\label{Fig:Fig1}
\end{figure}
%
%
\begin{figure}[h] 
\begin{center}
\includegraphics[scale=0.25]{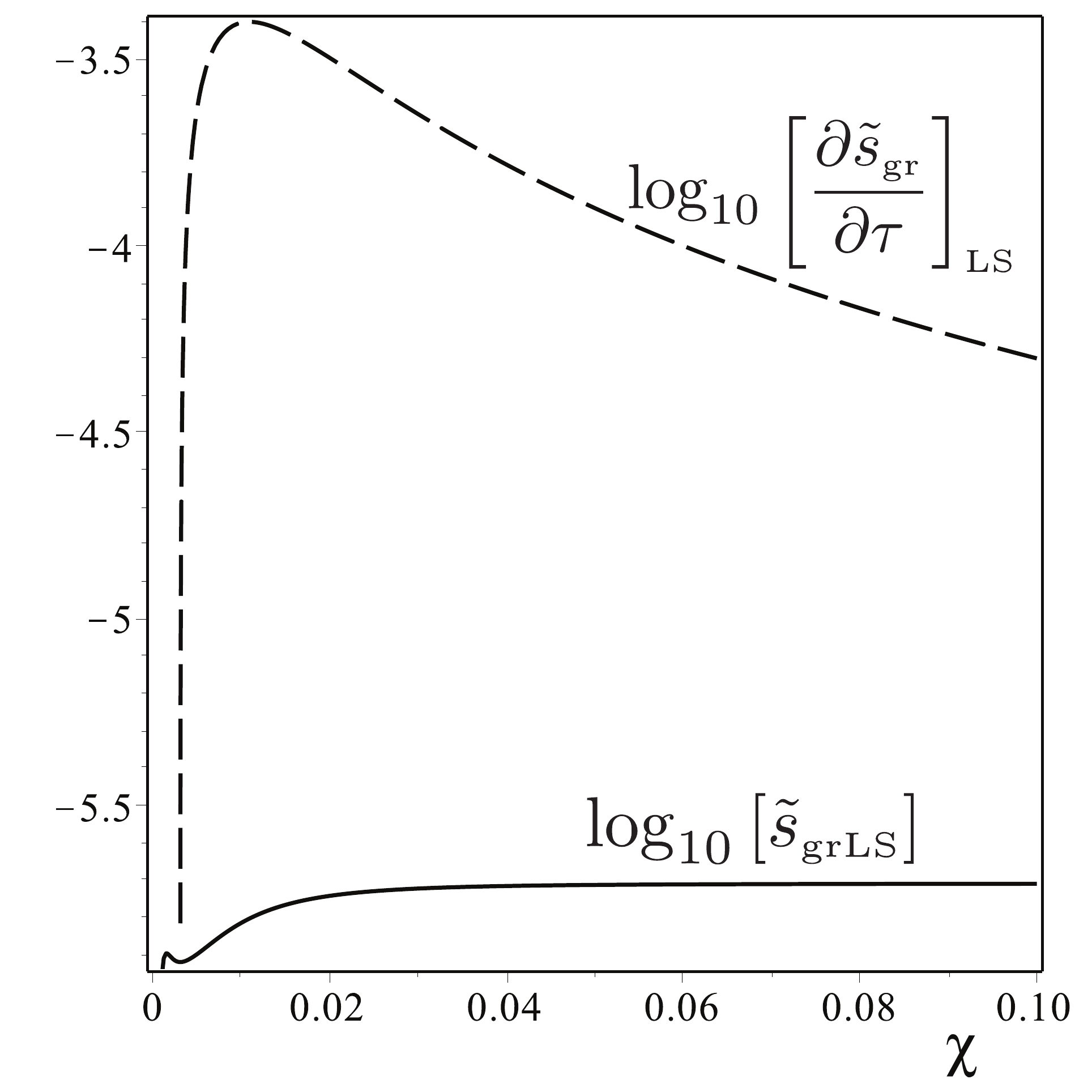}
\end{center}
\caption{{\bf Initial entropy state}. The figure displays the graph (solid curve) of the logarithm of the initial entropy state integral $\sgrls$ defined in (\ref{initstate}), and used to define the normalised CET entropy growth and its integral in (\ref{toplot}). For the purpose of comparison we included the logarithmic graph (broken curve) of the initial value of the time derivative $\partial\tilde\sgr/\partial\tau$. Notice that the latter is about three orders of magnitude larger than $\sgrls$.}
\label{Fig:Fig2}
\end{figure}
%
%
\begin{table}[h]
\begin{center}
\begin{tabular}{|c|c|c|c|}
\hline
 & CDM profile & Curvature profile & Present background \\
\hline
EdS & \begin{tabular}{c}$\Ommls=1.0$\\ $m_{1}=-0.005$\\ $m_{2}=0.001$\end{tabular} & \begin{tabular}{c}$\Omkls=0$\\ $k_{1}=-0.009$\\ $k_{2}=0.01$\end{tabular} & \begin{tabular}{c}$\OmmO=1$\\ $\bar H_{b0}=47\mbox{ km/s/Mpc}$\end{tabular} \\
\hline
OCDM & \begin{tabular}{c}$\Ommls=0.988250$\\ $m_{1}=-0.005$\\ $m_{2}=0.001$\end{tabular} & \begin{tabular}{c}$\Omkls=-0.00175$\\ $k_{1}=-0.009$\\ $k_{2}=0.01$\end{tabular} & \begin{tabular}{c}$\OmmO=0.27$\\ $\OmLO=0$\\ $\bar H_{b0}=59\mbox{ km/s/Mpc}$\end{tabular}\\
\hline
$\Lambda$CDM &   \begin{tabular}{c}$\Ommls=1.0$\\ $m_{1}=-0.005$\\ $m_{2}=0.001$\end{tabular} &  \begin{tabular}{c}$\Omkls=0$\\ $k_{1}=-0.009$\\ $k_{2}=0.01$\end{tabular}  & \begin{tabular}{c}$\OmmO=0.27$\\ $\OmLO=0.73$\\ $\bar H_{b0}=69\mbox{ km/s/Mpc}$\end{tabular}\\
\hline
\end{tabular}
\caption{{\bf Parameters for the various illustrative models}. These parameters define initial conditions at $t=\tls$ ($\tau=0$) (through the ansatz (\ref{initconds3})) to obtain the deep voids of $\sim 50$ Mpc extension at present day cosmic time $\tau=\tau_0$ whose characteristic features are displayed in Fig.~\ref{Fig:Fig1}. The numerical values were chosen to comply with $|\DDkls|\approx |\DDmuls|$ (or $|m_1|\approx |k_1|\sim 10^{-3}$). These are also the parameters used to define initial conditions in  Fig.~\ref{Fig:Fig3} and Fig.~\ref{Fig:Fig4}. Note that $\bar H_{b0}$ (background Hubble scalar at $t=t_0$) is not the local $\bar H_0$: \,it must be determined by demanding that the background complies with the cosmic age constraint $t_0\sim 13.7\, \hbox{Gyr}$ for values $\OmmO,\,\OmLO$ supported by observations.}
\label{Tab:Tab1}
\end{center}
\end{table}
For the initial conditions summarised in Table 1, the Big Bang time function $\tbb(r)$ is not simultaneous, having a background value $\bar\tbb=\tbb(\infty)=0$ that corresponds to $\tau=-2/3$ and a central value $\tbb(0)$ that corresponds to $\tau=-2/3+\delta_{\tbb}>-2/3$ (regularity conditions require $\tbb'\leq 0$), with $\delta_{\tbb}>0$ of the same order of magnitude as the maximal amplitude of the fluctuations: {\it i.e.} $\delta_{\tbb}\sim O(\hbox{max}(|m_1|,\,|k_1|))$. Hence, the Big Bang non--simultaneity is of the order of $10^{-3}\tls$ to $10^{-2}\tls$, or $\sim 10^2-10^3$ years, which is negligible in comparison with cosmic age \cite{susslar,part2}. Since LTB models are not valid for $t<\tls$ (or $\tau<0$) because of the non-negligible presence of radiation,  we only consider the evolution for $t>\tls$ (or $\tau>0$), and thus our numerical study is not affected by dominant decaying modes very close to the Big Bang time, as the latter are already subdominant at $\tls$ (see numerical examples in \cite{susslar,part2}).

\subsection{Numerical results.}\label{NUmerics3}

To examine numerically the evolution of the CET entropy for arbitrary fundamental observers in the desired void structures (see table 1 and Fig.~\ref{Fig:Fig1}), we need to evaluate numerically the initial entropy state at $\tls$ in (\ref{sgrLS}):  
\begin{equation}\fl \tilde\sgrls = \frac{\sgr(\chi)|_{\tau=0}}{\sigma} =-\frac{3}{2}\,\int_0^\chi{\muqls\frac{[\hqls+3\DDhls]_{,\chi}}{[h_{q,\tau}+3\DDH_{,\tau}]_{\textrm{\tiny{LS}}}}\times \frac{\DDhls\,\DDmuls}{|\hqls+3\DDhls|\,|\DDmuls|} \dd \chi},\label{initstate}\end{equation}
for these void structures. This integral is displayed in Fig.~\ref{Fig:Fig2} for the void in the $\Lambda$CDM background, together with $[\partial \tilde\sgr/\partial\tau]_{\textrm{\tiny{LS}}}$ (these plots are practically the same for voids in the other backgrounds).

The next step is to examine numerically the rate of growth in (\ref{sgrt2}) and the entropy integral in (\ref{sgrTOT}) normalised by the initial state (\ref{initstate}) 
\begin{equation}\fl \frac{\partial\SSgr}{\partial\tau},\quad \SSgr(\tau)|_\chi=\int_0^\tau{\left[\frac{\partial\SSgr}{\partial\tau}\right]_\chi\dd\tau},\qquad \SSgr(\tau)|_\chi\equiv 1+\frac{\tilde\sgr(\tau)|_\chi}{\tilde\sgrls}=1+\frac{\sgr(\tau)|_\chi}{\sgrls}.\label{toplot}\end{equation}
In Fig.~\ref{Fig:Fig3} we display the graph of the growth rate $\partial \SSgr/\partial\tau$ above as a function of $(\tau,\chi)$ for the void structures under consideration, while Fig.~\ref{Fig:Fig4} depicts this growth rate and its time integral $\SSgr(\tau)|_\chi$ for a typical fundamental observer ($\chi$ fixed) inside these voids. 
%
\begin{figure}[h] 
\begin{center}
\includegraphics[scale=0.50]{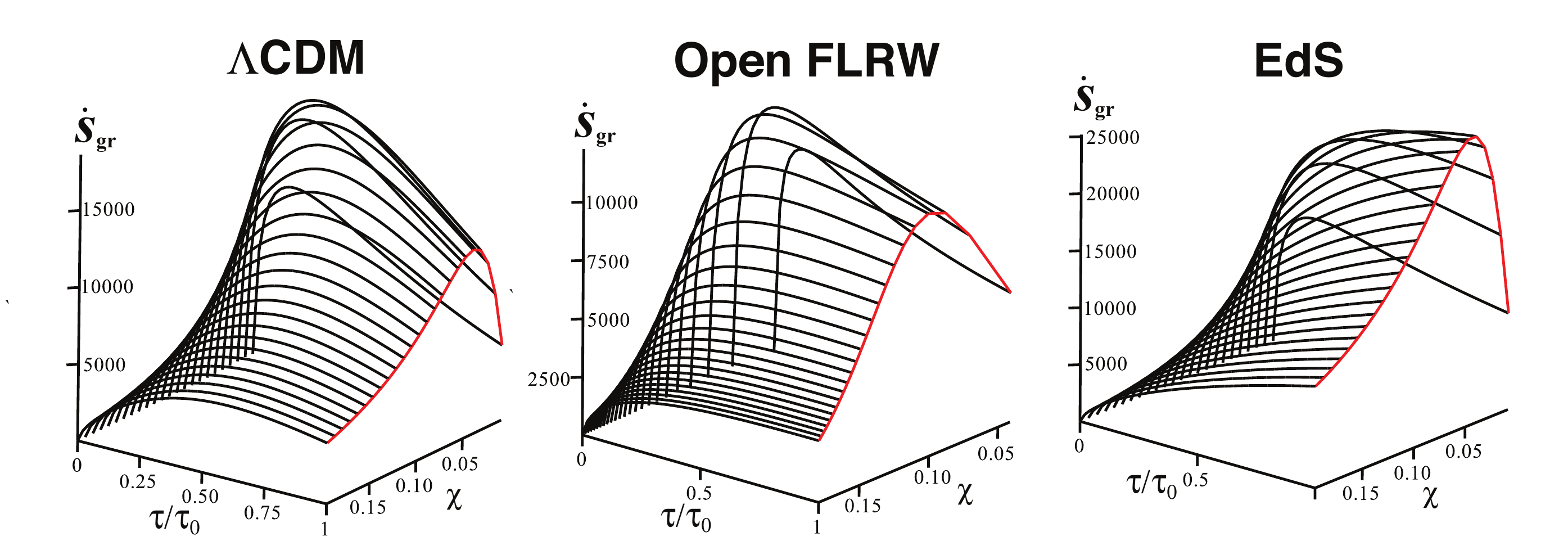}
\end{center}
\caption{Evolution of the rate of growth of the gravitational CET entropy $\partial \SSgr/\partial\tau$ as a function of $(\tau,\chi)$ for fundamental observers inside CDM voids in a $\Lambda$CDM background, an open FLRW background and an Einstein-de-Sitter (EdS) background. The parameters for each case are given in Tab.~\ref{Tab:Tab1}. The red curves correspond to present cosmic time $\tau=\tau_0$ (or $t=t_{0}=13.7.$ Gyr).  }
\label{Fig:Fig3}
\end{figure}
%
\begin{figure}
\begin{center}
\includegraphics[scale=0.45]{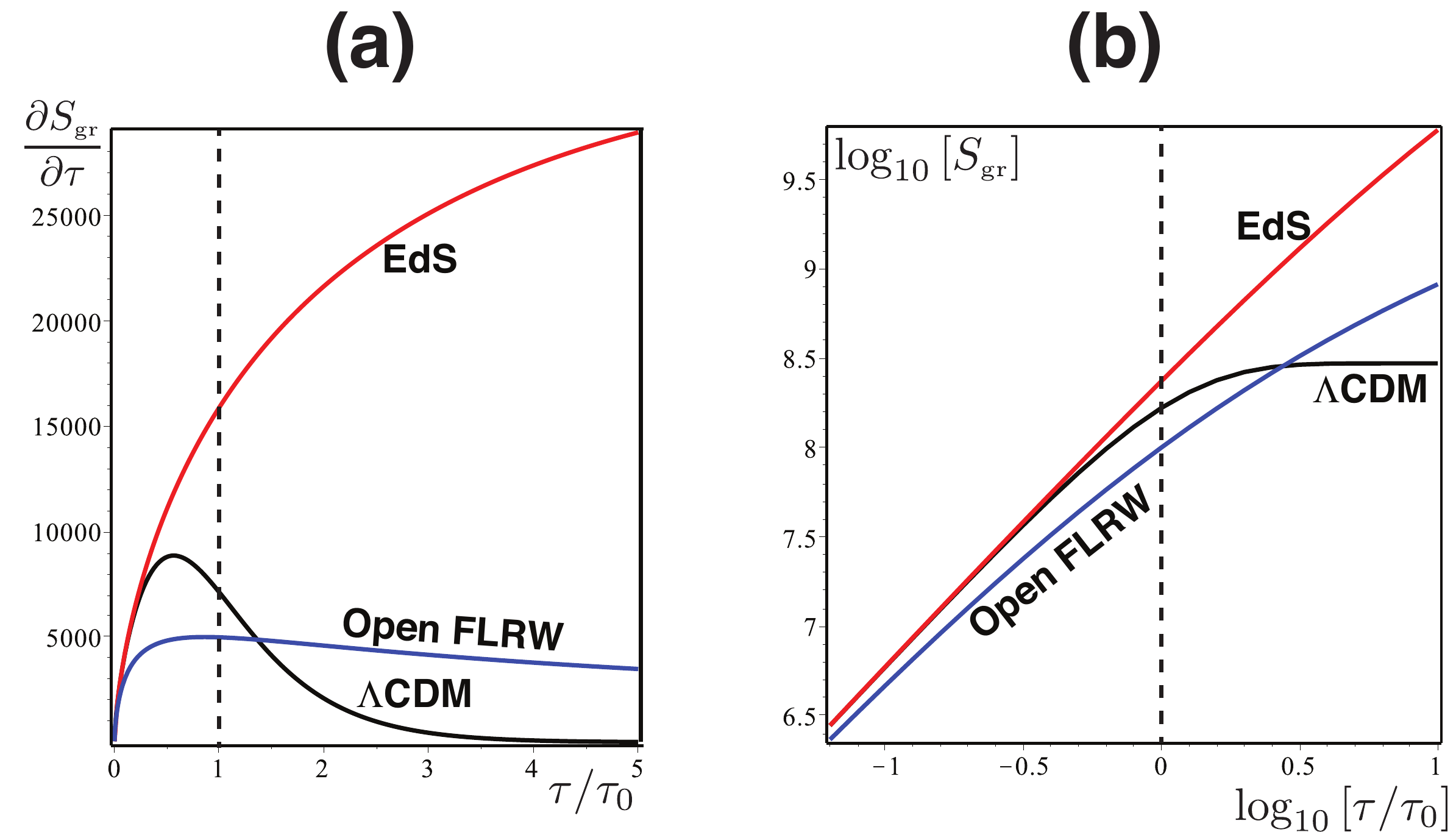}
\end{center}
\caption{{\bf Gravitational CET entropy for a typical comoving observer.} The curves correspond to a fixed comoving coordinate $\chi=0.15$ that represents an observer in the transition from the void centre to the background. Black, red and blue curves (online version) depict curves for voids in a $\Lambda$CDM, Einstein de Sitter and open FLRW backgrounds. Panel (a) displays the growth rate $\partial\SSgr/\partial\tau$, while panel (b) depicts $\log_{10}\SSgr$, both quantities defined in (\ref{toplot}). Notice how the growth rate starts decreasing well before present cosmic time $\tau_0$ for the voids in $\Lambda$CDM and open FLRW background (it starts decreasing at $\tau\sim 10\,\tau_0$ for the void with an EdS background). In panel (b) the logarithmic growth of $\SSgr$ is evident from the curves for the voids in the EdS and open FLRW backgrounds. Notice that at $\tau_0$ the CET entropy in the void in an open FLRW background is smaller than in the void in a $\Lambda$CDM background. However, for $\tau\sim 2\tau_0$ the CET entropy in the latter reaches a terminal value, while it keeps growing in the former.}
\label{Fig:Fig4}
\end{figure}
%
%
It is clear from Fig.~\ref{Fig:Fig3} and Fig.~\ref{Fig:Fig4} that the CET entropy increases for fundamental observers inside voids in the three backgrounds. The growth of $\partial\SSgr/\partial\tau$ and $\SSgr(\tau)|_\chi$ is more intense for the void in an EdS background ($\partial\SSgr/\partial\tau$ eventually starts decreasing for much larger times $\sim 10\,\tau_0$). This is an expected result from the asymptotic expansion (\ref{sgras2}), but can also be understood from the dependence of this growth rate on the Hubble scalar fluctuation $\DDH$, which (from (\ref{Dadef})) is closely connected with the radial gradients $\HH'_q$ and $\HH'$. Since $\HH_q$ and $\HH$ decay at a much slower rate for the EdS background and (at the same time) will take large values inside the void, these gradients are necessarily much larger for voids in this background than in those with a $\Lambda$CDM and open FLRW backgrounds. The larger growth of $\partial\SSgr/\partial\tau$ and $\SSgr$ for an EdS background is also reported in the figures of \cite{Marozzi}.    

It is straightforward to see in Fig.~\ref{Fig:Fig3} and in the panel (a) of Fig.~\ref{Fig:Fig4} that $\partial\SSgr/\partial\tau$ in the void in a $\Lambda$CDM background has already begun to decrease at $\tau_0$ (roughly as in figure 3 of \cite{Marozzi}).  Remarkably, as shown by the same figures, the negative curvature of the open FLRW background also leads to a suppression of the growth rate before $\tau_0$ (this case was not examined in \cite{Marozzi}). As expected from the asymptotic expansions (\ref{sgras1}) and (\ref{sgras2}), the graphs of Fig.~\ref{Fig:Fig3} and Fig.~\ref{Fig:Fig4}  confirm that $\partial\SSgr/\partial\tau\to 0$ holds for all voids, though at very different time rates. Therefore, as shown by panel (b) of Fig.~\ref{Fig:Fig4}, only for the void in a $\Lambda$CDM background the exponential decay is sufficiently rapid to allow for $\SSgr$ to reach terminal ``saturation'' values at (roughly) $\tau\sim 2\tau_0$ for all fundamental observers. As shown in  Fig.~\ref{Fig:Fig6}, the saturation values of $\SSgr$ are position dependent inside the void, ranging between $10^8$ to almost $10^9$ times its initial state $\sgrls$. 
  
For the void in an open FLRW background the logarithmic decay of $\partial\SSgr/\partial\tau\to 0$ leads to an asymptotic logarithmic growth of $\SSgr$ (panel (b) of Fig.~\ref{Fig:Fig4}). Hence, negative spatial curvature does yield a stronger suppression of the entropy rate of growth in comparison with the spatially flat EdS background, but it is not sufficiently strong to yield a finite terminal value for $\SSgr$. The time profile of the curves of panel (b) of Fig.~\ref{Fig:Fig3} for the $\Lambda$CDM and EdS backgrounds are qualitatively analogous to the first and fourth curves (top and bottom) of figure 2 of \cite{Marozzi} that correspond to these backgrounds. 

\section{Comparison with the perturbative treatment.}\label{Comparison}

It is important to compare the results obtained in previous sections with those of the perturbative analysis in \cite{Marozzi}. If we bear in mind the correspondence between fluctuations of the q--scalars and spherical dust perturbations of the linear theory (see detail in \cite{SHDG}), the common theoretical ground (related to the Weyl tensor) between the non--perturbative and perturbative results can be appreciated by a direct comparison between the conditions for entropy growth in both treatments. Since the Weyl tensor and its squared contraction for LTB models take the form \cite{susslar}
\begin{equation}\fl C^{ab}_{cd}=-\frac{4\pi}{3}\DDrho\,\left(h^{[a}_{[c}-3u_{[c}u^{[a}\right)\,\eigf^{b]}_{d]},\qquad \sqrt{C_{abcd}C^{abcd}} =\frac{16\pi}{\sqrt{3}}|\DDrho|,\end{equation}
it is then evident that equation (\ref{CETc12}) (and similar equations in \cite{susslar,sussAN}) and equation (37) of \cite{Marozzi} are equivalent at first order 
\begin{equation}\frac{\partial}{\partial t}\left(|\DDrho|\,a^3\,\Gamma\right)\sim \frac{\bar a}{\bar\HH}\frac{d}{d\eta}\left(\bar a^3\sqrt{\frac{\langle \overline{C_{abcd}C^{abcd}}\rangle_{\cal D}}{192}}\right),\end{equation}
where $\bar a$ and $\bar\HH$ are the FLRW scale factor and Hubble scalar, $\eta=\int{\dd t/\bar a}$ is the conformal time and we used the fact that at linear first order we must have (see \cite{SHDG}): $a\sim \bar a,\,\,\Gamma\sim 1$, as well as $\langle \overline{C_{abcd}C^{abcd}}\rangle_{\cal D}\sim C_{abcd}C^{abcd}\propto |\DDrho|^2$, where the overline in $\langle \,\,\rangle_{\cal D}$ denotes an ensemble average (see \cite{Marozzi}).

The connection between the perturbative and non--perturbative approach to the CET entropy growth can also be appreciated from the relation between the exact LTB evolution equations and evolution equations for linear perturbations. To highlight this relation we recast the system (\ref{FFq1})--(\ref{FFq4}) into a dimensional form that is suitable for comparison with evolution equations for dust perturbations in a synchronous comoving gauge (see comprehensive discussion in \cite{SHDG}):
\bse\ba\fl \dot{\bar\rho}_{\textrm{\tiny{as}}} =-3\rhobaras\,\HHbaras,\label{asev1}\\
\fl\dot{\bar\HH}_{\textrm{\tiny{as}}}=-\HHbaras^2-\frac{4\pi}{3}\rhobaras+\frac{8\pi}{3}\Lambda,\label{asev2}\\
\fl\dot\Drhoas = -3(1+\Drhoas)\DDHas,\label{asev3}\\
\fl\dot\DDHas = -\left(6\HHbaras-4\HH_q+3\DDHas\right)\DDHas-2(\HH_q-\HHbaras)^2-\frac{4\pi}{3}\rhobaras\,\Drhoas,\label{asev4}\ea\ese  
where $\rhobaras(t),\,\HHbaras(t)$ are the limits of $\rho,\,\HH$ (and also of $\rho_q,\,\HH_q$) as $r\to\infty$ for arbitrary fixed $t$, and thus correspond to the density and Hubble scalar of the asymptotic FLRW background (in fact, (\ref{asev1})--(\ref{asev2}) determine this background), while $\Drhoas,\,\DDHas$ are the asymptotic non--local fluctuations  defined as 
\begin{equation}\Drhoas = \frac{\rho-\rhobaras}{\rhobaras},\qquad \DDHas = \HH-\HHbaras.\label{asflucs}\end{equation}
whose direct connection to the gauge invariant density and Hubble scalar perturbations of linear perturbation theory is discussed extensively in \cite{SHDG} (in particular, $\Drhoas$ also corresponds to the density contrast).   

A linear regime associated with (\ref{asev1})--(\ref{asev4}) and (\ref{asflucs}) can be defined by the conditions $|\Drhoas|\ll 1$ and $|\DDHas/\HHbaras|\ll 1$, which imply (as long as they remain valid) that the quadratic terms $\Drhoas\DDHas$ and $[\DDHas]^2$ are negligible, and thus $\HH\approx\HH_q\approx \HHbaras$ hold. Hence, the dynamics of the dust fluctuation (in this regime) is determined by the linearised forms of (\ref{asev3})--(\ref{asev4}):
\begin{equation} \fl\dot\Drhoas = -3\DDHas,\qquad  \dot\DDHas = -2\HHbaras\DDHas-\frac{4\pi}{3}\rhobaras\,\Drhoas,\label{asevlin}\end{equation}
which, as shown in \cite{SHDG}, are fully equivalent to the evolution equations of standard dust cosmological perturbations in the isochronous comoving gauge. In fact, if combined (\ref{asevlin}) lead to the well known second order equation $\ddot \Drhoas+2\HHbaras\dot\Drhoas-4\pi\rhobaras\Drhoas=0$ for linear dust perturbations. 

Since void structures defined by the initial conditions at $\tls$ listed in Table 1 evolve into a fully non--linear regime at $\tau_0$ (see Fig.~\ref{Fig:Fig1}), it is straightforward to show that these structures necessarily comply with a linear regime  for restricted evolution times that are close to the initial time $\tls$ (or $\tau\approx 0$). Hence, we can state unequivocally that the entropy growth rate, $\partial \SSgr/\partial\tau$, and the entropy integral, $\SSgr(\tau)|_\chi$, reach much smaller values in the linear regime than in the non--linear regime. This becomes evident by comparing the early times ($\tau\approx 0$) linear regime values of these quantities with their non--linear regime latter times ($\tau>\tau_0/2$) in their plots of Fig.~\ref{Fig:Fig3} and Fig.~\ref{Fig:Fig4}.  

However, a more accurate quantitative comparison between the linear and non--linear regime can be achieved by looking at the evolution of the CET entropy for void structures that remain in the linear regime up to $\tau_0$. From the discussion of section \ref{Numerics2}, it is reasonable to guess that the right initial conditions for such voids involve choosing values of the amplitude $|k_1|$ of the initial spatial curvature fluctuation $\DDkls$ that are smaller than those of Table 1, since  this amplitude is the determinant parameter for the rate of growth of the CET entropy.     
  
Considering now only voids in a $\Lambda$CDM background, we build up initial conditions as described in the previous paragraph:  fluctuations complying with $|\DDKKls| \ll |\DDrhols|\sim O(10^{-3})$, which can be achieved by the ansatz (\ref{initconds3}) with $|k_1|< |m_1|=0.001$. For such initial data we solve numerically the linear evolution equations (\ref{asev1}), (\ref{asev4}) and (\ref{asevlin}). The density contrast and Hubble scalar of the resulting void configurations are displayed by Fig.~\ref{Fig:Fig5}, where the curves marked by B, C and D correspond to the following decreasing values: $k_1=-9\times 10^{-4},\,-9\times 10^{-5},\,-9\times 10^{-6}$, with the parameters $m_1,\,m_2,\,k_2$ kept the same as the void in the $\Lambda$CDM background in Table 1. For the purpose of comparison, we include the curves marked by A that correspond to the non--linear void in a $\Lambda$CDM background used in Figs.~\ref{Fig:Fig1}, \ref{Fig:Fig3} and ~\ref{Fig:Fig4}. As shown by the curves B, C and D of Fig.~\ref{Fig:Fig5}, the voids remain linear ({\it i.e.} very ``shallow'') up to present cosmic time $\tau_0$.

In order to compare the CET entropy growth of the voids in the linear and non--linear regimes, we plotted in Fig.~\ref{Fig:Fig6} the saturation terminal value of $\SSgr$ (panel (a)) and the saturation per comoving volume (panel (b)) for fundamental observers in the voids of Fig.~\ref{Fig:Fig5}. Since $\SSgr$ reaches its saturation at $\tau\sim 2\tau_0$, the saturation values are obtained by evaluating $\SSgr$ at $\tau=3\tau_0$, whereas the ratio of saturation to comoving volume is obtained by plotting $\SSgr(3\tau_0)/R^3(3\tau_0)$, where $R=a\,r$ is the area distance. As expected, the CET entropy growth in linear voids is much smaller than in non--linear ones, as the magnitudes of $\SSgr(3\tau_0)$ and $\SSgr(3\tau_0)/R^3(3\tau_0)$ for the curves B, C and D in Fig.~\ref{Fig:Fig6} decrease (with respect to the curve A) in direct proportionality to the decrease of the amplitude parameter $|k_1|$. 

In particular, it is important to emphasise that the numerical results of panel (b) can be directly compared to the numerical results of \cite{Marozzi}, who only examined saturation CET entropy values per comoving volume. Notice how the curves B, C and D in panel (b) of Fig.~\ref{Fig:Fig6} for the linear voids rapidly decrease to values of $\sim 10^{-4}$, which is the numerical value displayed in the fourth curve (top to bottom) of figure 2 of \cite{Marozzi} for the CET saturation per comoving volume in linear perturbations in a $\Lambda$CDM background. Notice that also in the non--linear void (curve A) the saturation values per volume decrease to the perturbative values $\sim 10^{-4}$ for increasing comoving volume,    as the non--linear void becomes linear at large scales where its parameters approach the background parameters.   

As a test of consistency, we verified that exactly the same curves of Fig.~\ref{Fig:Fig5} and Fig.~\ref{Fig:Fig6} are obtained (for same initial conditions) from the full exact evolution equations (either (\ref{FFq1})--(\ref{FFq4}) or (\ref{asev1})--(\ref{asev4})). We also verified (but are not displaying the graphs) that $\SSgr$ diverges logarithmically for linear voids in an open FLRW and Einstein de Sitter backgrounds. Hence, our numerical results fully agree with those of \cite{Marozzi}. 
%
\begin{figure}
\begin{center}
\includegraphics[scale=0.40]{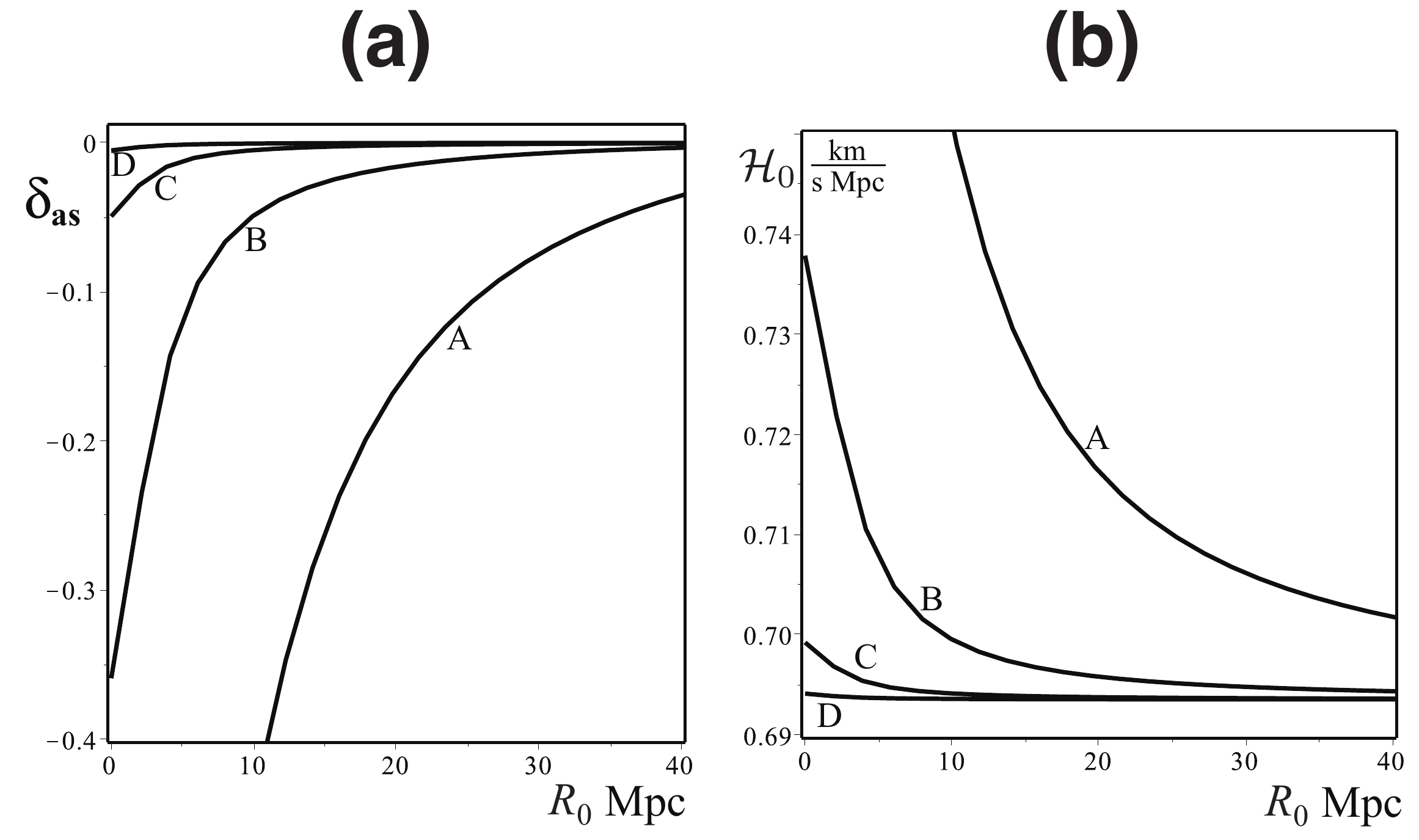}
\end{center}
\caption{{\bf Voids in the linear regime.} The figure depicts the same quantities of panels (a) and (b) of Fig.~\ref{Fig:Fig1} (density contrast and Hubble scalar), but for voids in a linear regime and in a $\Lambda$CDM background (the curve marked by A corresponds to the non--linear $\Lambda$CDM void used in previous figures). The curves marked by B, C and D correspond to the linear voids whose parameters are defined in the text. Notice that the voids associated with the curves B, C and D have the negligible density contrast and radial variation of $\HH_0$ expected from the linear regime.}
\label{Fig:Fig5}
\end{figure}
%
%
\begin{figure}
\begin{center}
\includegraphics[scale=0.40]{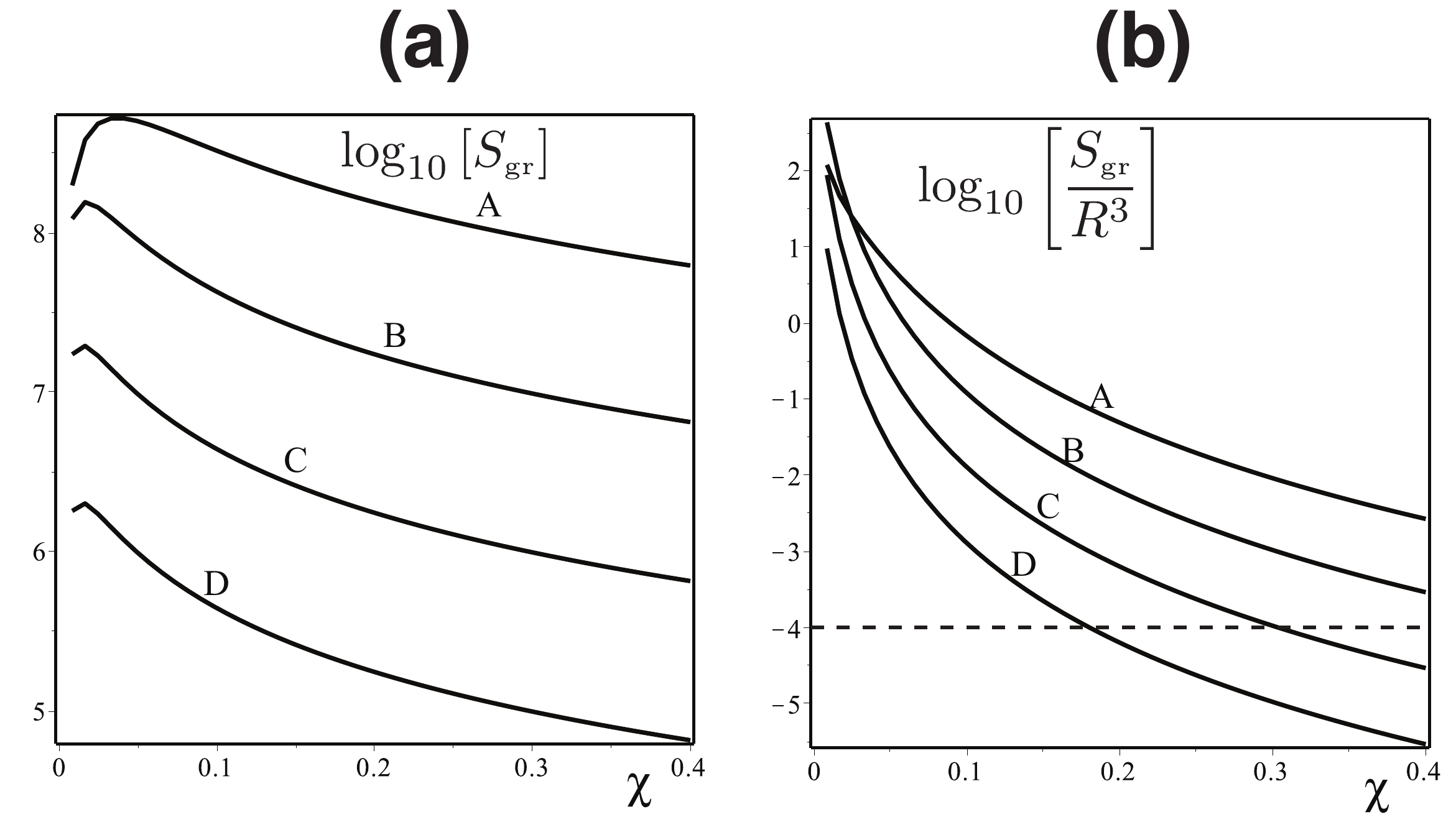}
\end{center}
\caption{{\bf Saturation values.} Panel (a) depicts the logarithm of the terminal CET entropy saturation values ($\SSgr$ at $\tau=3\tau_0$) for fundamental observers in the voids of Fig.~\ref{Fig:Fig5}. Panel (b) displays the same saturation values weighed by their comoving volume defined by $R^3$, where $R=a\,r$ is the area distance at $3\tau_0$. Notice that the saturation values decrease several orders of magnitude for the linear regime voids (curves B, C and D). On the other hand, the saturation value per volume reaches for the linear voids (and the non--linear one at large scales) the same order of magnitude numerical values $\sim 10^{-4}$ reported by in Fig 2 of the perturbative study of \cite{Marozzi}. }
\label{Fig:Fig6}
\end{figure}

\section{Conclusion.}\label{Conclusion} 

We have examined the exact non--perturbative effect of negative curvature and $\Lambda$ term on the evolution of the CET gravitational entropy, $\sgr$, inside of expanding sub--horizon 50-100 Mpc voids that form in three prototypical standard FLRW cosmological backgrounds: $\Lambda$CDM concordance-type model, open FLRW (negative spatial curvature but $\Lambda=0$) and Einstein de Sitter (EdS) (see Table 1 and Fig.~\ref{Fig:Fig1}). We used analytic expressions to examine the asymptotic time behaviour of $\dot\sgr$ and its line integral $\sgr(t)|_r$, and a numerical study to determine their full evolution from linear initial conditions specified at the last scattering time $t=\tls$.  

We have found that for all such voids the rate of entropy production $\dot\sgr$ is initially growing, but this growth has started to be suppressed before our cosmic time $t=t_0$ in voids in the $\Lambda$CDM and open FLRW backgrounds, while for voids in an EdS background this suppression only begins at very late times $t\sim 10\,t_0$. While for all voids we have $\dot\sgr\to 0$ asymptotically, this decay occurs at a rapid exponential rate for voids in a $\Lambda$CDM background and it occurs at a slow logarithmic pace for voids whose FLRW background has $\Lambda=0$ (open FLRW and EdS). As a consequence, only for voids in a $\Lambda$CDM background this asymptotic decay is sufficiently rapid to allow $\sgr(t)|_r$ to converge asymptotically for all fundamental observers (at about $t\sim 2\,t_0$) into a terminal equilibrium (or ``saturation'') value between $10^8$ and $10^9$ times its initial state defined by the radial entropy integral (\ref{initstate}) at $t=\tls$ fixed (see Fig.~\ref{Fig:Fig2}). For voids in the open FLRW and EdS backgrounds $\sgr(t)|_r$ diverges logarithmically (though at a much slower pace for the former). These results are all depicted graphically by Fig.~\ref{Fig:Fig3} and Fig.~\ref{Fig:Fig4}.

Since our results should agree in the appropriate regime with those of \cite{Marozzi} based on linear dust perturbations, we provided a direct comparison between the exact non--linear evolution of the CET entropy for voids (in an $\Lambda$CDM background) that are fully non--linear at $t_0$ and voids (whose initial conditions are defined also at $\tls$) that remain in a linear regime up to $t_0$ (see Fig.~\ref{Fig:Fig5}). The results of this comparison are as expected: as shown in Fig.~\ref{Fig:Fig6} the saturation terminal values of the CET entropy decrease for the linear voids in the same order of magnitude proportion as the amplitude of the initial spatial curvature fluctuation.
However, we also obtained a good quantitative agreement with the results of \cite{Marozzi}: the saturation values of the CET entropy per comoving volume (the curves B, C and D in panel (b) of Fig.~\ref{Fig:Fig6}) rapidly decay to values of $\sim 10^{-4}$ that fit nicely with the same quantity plotted in curves in Fig 2 of \cite{Marozzi} (see specially the fourth curve from top to bottom that corresponds to the $\Lambda$CDM background). In fact, this agreement also holds for the non--linear void (curve A), which shows that linear perturbative conditions should also hold in the large scales for non--linear structures where the latter converge to the FLRW background.                     
       
It is worth mentioning from a theoretical perspective that the asymptotic value of the CET gravitational temperature for voids in a $\Lambda$CDM background is proportional to the cosmological constant (this temperature is zero for voids in $\Lambda=0$ backgrounds). This fact indicates the existence of a potential link between the value of the cosmological constant and thermodynamical considerations, supporting the notion that gravity and thermodynamics might be intertwined at a fundamental level; see e.g. \cite{paddy} for attempts at such an emergent gravity scenario.

Another important theoretical issue is the fundamental nature of the gravitational entropy and its connection to other entropy concepts in the context of gravitational systems, such as the holographic Hawking--Bekenstein black hole entropy \cite{HawBek,PadRevs} and its various extensions to fluid sources (in FLRW spacetimes \cite{HoloFLRW} and in LTB models \cite{bolstoeg,misra,holoLTB}). Since this issue is still an open problem whose proper treatment would require a separate article to be submitted in future work, we introduced in section \ref{DimVars} a fiducial fundamental constant (or scale parameter) with entropy units, $\sigma$, in order to examine the evolution of the CET entropy through dimensionless expressions. As a consequence, the numerical values we have obtained for the CET entropy need to be understood as entropy values proportional to this yet unknown fundamental entropy. See \cite{Rituetal} for a recent way of relating the Hawking-Bekenstein entropy of a black hole to the variation of the CET entropy outside the collapsing dust at the origin of the black hole.

Evidently, we have only examined very idealized spherical expanding cosmic voids, and thus further research is needed to probe the CET proposal (and the proposal of \cite{HB}) on more general spacetimes, such as Szekeres models \cite{sussbol}, and on the process of structure formation and gravitational collapse. In particular, we aim at studying the growth of these gravitational entropies in the context of the formation of virialized stationary structures, which may provide a connection with theoretical work done on n--body numerical simulations \cite{nbody} and Newtonian self--gravitational systems \cite{Newtonian}, as well as research on various proposals on non--extensive entropy definitions \cite{Tsallis}. This research is currently under way and will be the subject of future works.


\section*{Acknowledgments:}
RS acknowledges financial support from grant PAPIIT-DGAPA IA101414 and SEP-CONACYT 239639.\
JL's work is supported by the National Research Foundation (South Africa).


\end{document}